\documentclass[a4paper,12pt,twoside]{article}
\usepackage[british]{babel}
\usepackage{amssymb,amsmath,amsfonts,verbatim,xkeyval,bm}
\usepackage{graphicx}
\usepackage{subfigure}
\newcommand{\ds}{\displaystyle}
\def\Lie{{L}}

\let\golden=\phi

\newcommand{\pluseq}{\hookleftarrow}
\def\vet#1{{\underline #1}}
\def\build#1_#2^#3{\mathrel{
\mathop{\kern 0pt#1}\limits_{#2}^{#3}}}
\def\reali{\mathbb{R}}
\def\complessi{\mathbb{C}}
\def\naturali{\mathbb{N}}
\def\interi{\mathbb{Z}}
\def\toro{\mathbb{T}}

\def\diff{{\rm d}}
\def\Ascr{\mathcal{A}}
\def\Bscr{\mathcal{B}}
\def\Cscr{\mathcal{C}}
\def\Dscr{\mathcal{D}}

\def\Gscr{\mathcal{G}}
\def\Hscr{\mathcal{H}}
\def\Jscr{\mathcal{J}}
\def\Kscr{\mathcal{K}}

\def\Oscr{\mathcal{O}}
\def\Pscr{\mathcal{P}}
\def\Rscr{\mathcal{R}}
\def\Sscr{\mathcal{S}}

\def\Vscr{\mathcal{V}}
\def\Zscr{\mathcal{Z}}

\def\epsilon{\varepsilon}
\def\rho{\varrho}
\def\phi{\varphi}
\def\csi{\xi}
\def\imunit{{\rm i}}
\def\poisson#1#2{\lbrace #1,#2 \rbrace}

\def\Hamvecfield{\Vscr_{\Hscr}}

\def\Pgot{{\mathfrak P}}

\title{\bf Quasi-periodic motions in a special\\ class of dynamical
  equations\\ with dissipative effects:\\ a pair of detection
  methods}

\author{
{\bf UGO LOCATELLI}\\
{\small Dipartimento di Matematica, 
Universit\`a degli Studi di Roma ``Tor Vergata'',}\\
{\small Via della Ricerca Scientifica 1, 00133--Roma (Italy).}\\
{\bf LETIZIA STEFANELLI}\\
{\small Geoazur, Universit\'e de Nice Sophia-Antipolis, Observatoire de la C\^ote d'Azur}\\
{\small 250, rue Albert Einstein, 06560  Valbonne (France).}\\
{\small e-mails:
  {\tt locatell@mat.uniroma2.it, stefanel@oca.eu}}
}

\date{}

\begin{document}
\maketitle

\selectlanguage{british}
\thispagestyle{empty}

\begin{abstract}
We consider a particular class of equations of motion, generalizing to
$n$ degrees of freedom the ``dissipative spin--orbit problem'',
commonly studied in Celestial Mechanics. Those equations are
formulated in a pseudo-Hamiltonian framework with action-angle
coordinates; they contain a quasi-integrable conservative part and
friction terms, assumed to be linear and isotropic with respect to the
action variables. In such a context, we transfer two methods
determining quasi-periodic solutions, which were originally designed
to analyze purely Hamiltonian quasi-integrable problems.

First, we show how the frequency map analysis can be adapted to this
kind of dissipative models. Our approach is based on a key
remark: the method can work as usual, by studying the behavior of the
angular velocities of the motions as a function of the so called
``external frequencies'', instead of the actions.

Moreover, we explicitly implement the Kolmogorov's normalization
algorithm for the dissipative systems considered here. In a previous
article, we proved a theoretical result: such a constructing
procedure is convergent under the hypotheses usually assumed in KAM
theory.  In the present work, we show that it can be translated to a
code making algebraic manipulations on a computer, so to calculate
effectively quasi-periodic solutions on invariant tori.

Both the methods are carefully tested, by checking that their
predictions are in agreement, in the case of the so called ``dissipative forced
pendulum''. Furthermore, the results obtained by applying
our adaptation of the frequency analysis method to the dissipative
standard map are compared with some existing ones in the literature.
\end{abstract}

\bigskip

\markboth{U. Locatelli, L. Stefanelli}{Quasi-periodic motions in $\ldots$
  equations with dissipative effects $\ldots$ detection methods}

\section{Introduction}\label{sec:intro}

Why the Moon shows us always the same side? This is one of the most
ancient scientific questions raised by the observation of the sky.
The data made available by modern spatial missions clearly showed that
the spin-orbit periodic motion is a rather common phenomenon in our
solar system. Here, a $p$:$q$ spin-orbit resonance means that the
satellite turns on its spin axis $p$ times while doing $q$ revolutions
around its star/planet. Actually, more than 20 planet--satellite pairs
have been observed to stay into the 1:1 spin-orbit resonant state,
while just one planet (Mercury) shows a different periodic behavior,
because it rotates three times on itself during two complete
revolutions around the Sun. A convincing explanation of the capture in
resonance for the case of Mercury is provided in~\cite{Cor-Las-2004},
where its present state is explained as a consequence of the fact that
in the past the Mercury's orbit was much more eccentric. This scenario
is discussed within the framework of a spin-orbit model including a
dissipative force depending linearly on the relative angular velocity
(for its introduction see also~\cite{Gol-Pea-1966},
\cite{Gol-Pea-1970}, \cite{MacDonald-1964} and~\cite{Peale-2005}).  In
this model, different periodic orbits can coexist and the measure of
their basins of attraction can be evaluated both in a numerical and in
an analytic way (see~\cite{Cel-Chi-2008} and~\cite{Bia-Chi-2009}).
Within the different context of a viscoelastic model of the satellite,
it has been recently shown that the capture into the 1:1 spin-orbit
resonance is the generic final fate of such a dissipative system
(see~\cite{Bam-Haus-2012} and~\cite{Bam-Haus-2014}).

In the last few years, the data about the rotational motion of some
planets and satellites (e.g., Mercury, Titan and Europa) has been
related to the study of their internal structure; this renewed the
interest in the rotational dynamics of a non-rigid celestial body. In
this context, an important role is played also by small oscillations
around periodic orbits, which are also due to the perturbations
exerted by other planets (see, e.g.,~\cite{DHo-Lem-2008}).  Therefore,
more and more sophisticated numerical tools are required to analyze
this kind of weakly-dissipative systems.

In the present work, we adapt a numerical method and a semi-analytic
one usually devoted to the study of Hamiltonian systems, in order
to improve the description of the invariant attractors in the
dissipative framework. The first one is the frequency map analysis
and the semi-analitical one is the constructive algorithm of the Kolmogorov's normal form.
\smallskip

The frequency map analysis has been originally designed by J.~Laskar
to study conservative systems (see~\cite{Laskar-99}
and~\cite{Laskar-2005} for an introduction, while,
e.g.,~\cite{Gom-Mon-Sim-2010.1}, \cite{Gom-Mon-Sim-2010.2}
and~\cite{Noullez-2008} are devoted to interesting alternative
approaches).  It is a powerful tool used to investigate the chaotic
regions and those filled by invariant tori in several Hamiltonian
systems (see, e.g.,~\cite{Cha-Las-Ben-Jau-2001}, \cite{Dum-Las-1993},
\cite{Las-Rob-1993}, and~\cite{Pap-Las-1998}) as well as in symplectic
mappings (see~\cite{LFC}). In particular, the study of the variation
of the fundamental frequencies allows to make a detailed cartography
of the regular and chaotic regions in the Solar System
(see~\cite{Rob-Las-2001}).

In the present work, we mainly focus on the so called
{\sl dissipative forced pendulum}; the Newton equation for this model can
be written in the following form:
\begin{equation}
\ddot x+\eta(\dot x-\Omega)+\epsilon\frac{\partial U}{\partial x}(x,t)= 0
\ ,
\label{eq:pendolo-dissip-model-Newt}
\end{equation}
where $x\in\toro$ is an angle, $\epsilon$ is a (small) parameter and
the potential $U$ depends periodically both on $x$ and the time $t\,$.
Let us highlight the peculiar structure of the friction term
$\eta(\dot x-\Omega)$ appearing
in~(\ref{eq:pendolo-dissip-model-Newt}): it is linearly depending on
the momentum $\dot x$ and it contains the so called external frequency
parameter $\Omega\,$. The Newton equations of both the {\sl
  dissipative spin-orbit model} and the {\sl dissipative forced
  pendulum} are of type~(\ref{eq:pendolo-dissip-model-Newt}), but the
numerical explorations of the latter system require less computational
resources than those needed by the former one. By the way, let us
recall that the KAM--like theorems described in~\cite{Cel-Chi-2009}
and~\cite{Stef-Loc-2012} apply to models described by the
equation~(\ref{eq:pendolo-dissip-model-Newt}); moreover, dynamical
systems including dissipative terms (which, of course, are not
Hamiltonian) have been extensively studied in the last decades (see,
e.g.,~\cite{Bro-Hui-Sev-1996} and~\cite{Bro-Simo-Tat-1998}).

Our numerical approach is based on the study of the regularity of the
map $\Omega\mapsto\omega_1(\Omega)\,$, where the frequency $\omega_1$
is related to the eventually existing quasi-periodic solution
$t\mapsto x(\omega_1t\,,\,t)$ of
equation~(\ref{eq:pendolo-dissip-model-Newt}). We will show that our
investigation method is very similar to the one focusing on the
action-frequency map, which is commonly used for conservative
systems; moreover, our approach applies also to
dissipative mappings. In that context, a different frequency analysis
has already been used in~\cite{Ce-Fro-Le06}, for the study of the map
relating the frequency $\omega_1$ to the dissipation coefficient
$\eta\,$, for various fixed values of the perturbing parameter
$\epsilon\,$.

One of the issues of our numerical method concerns the determination
of the breakdown threshold (with respect to the small parameter ruling
the size of the perturbing terms) of the invariant tori. This will
allow us to compare our results with those given by other techniques,
which have been widely tested in the literature.  Among the known
methods, Greene's technique (for an introduction, see~\cite{Greene79}
and~\cite{MacKay-92}) provides the smallest uncertainity on the value
of the breakdown threshold for {\it symplectic} mappings, but it is
not so effective when dissipative terms are taken into account. This
is due to the fact that the Greene's method is based on the
calculation of a quantity (usually called residue), that is related to
the eigenvalues of the monodromy matrix associated to a full cycle of
a periodic orbit. Unfortunately, when the dissipation is introduced,
each periodic orbit of fixed frequency $\omega_1$ exists if and only
if the external frequency parameter
$\Omega\in[\Omega_{\omega_1;\,-},\Omega_{\omega_1;\,+}]\,$; moreover,
when the order of resonance related to $\omega_1$ is increased, the
interval $[\Omega_{\omega_1;\,-},\Omega_{\omega_1;\,+}]$ gets smaller
and smaller. Considering the mid value of the interval
$[\Omega_{\omega_1;\,-},\Omega_{\omega_1;\,+}]$ is a good way to adapt
the Greene's method to the dissipative standard map (as explained
in~\cite{Cal-Cel-2010} and~\cite{Cel-Diru-2011}), but this interval is
more and more difficult to locate for high order resonances; this
limits the strength of the method.  Another (recently established)
technique evaluates the breakdown threshold, by studying the Sobolev
norms of the function parametrizing the solution
(see~\cite{Cal-dlL-2010}). This approach apparently does not suffer
any particular drawback, when dissipative terms are taken into
account; therefore, it is able to determine the breakdown threshold
with many significant digits (see~\cite{Cal-Cel-2010}). Thus, the
comparison with those results on dissipative mappings will represent a
challenging test for our numerical method.
\smallskip

The semi-analytic method developed in the present paper strictly
concerns with KAM theory adapted to dissipative systems.  It is well
known that the original versions of the KAM theorem ensure the
existence of invariant tori filled by quasi-periodic orbits, in the
context of both Hamiltonian systems and symplectic mappings, which are
slightly perturbed with respect to some  integrable approximations
(see~\cite{Kolmogorov-1954}, \cite{Moser-1962}
and~\cite{Arnold-1963.1}). In his first and last article on KAM
theory, Kolmogorov pointed Celestial Mechanics as a field where such
result could be naturally applied; his vision was definitely fruitful
(see, e.g.,~\cite{Leontovich-62}, \cite{Dep-DepBar-67},
\cite{Celletti-90}, \cite{Celletti-90.1} and~\cite{Cel-Chi-2007}).
Actually, his proof scheme is based on the construction of sequences
of canonical transformations and of the corresponding Hamiltonians,
which are proved to converge (under suitable hypotheses) to the
so called Kolmogorov's normal form (see~\cite{Ben-Gal-Gio-Str-1984}
and~\cite{Chierchia-2008}). In~\cite{Gio-Loc-1997.1}, it is proved
that such a constructive algorithm can be rewritten according to a
{\it classical} scheme (being $\Oscr(\epsilon^{r})$ the size of the
generating function of the $r$--th canonical transformation), in such
a way to avoid the original {\it quadratic} convergence analogous to
the Newton method (where the generating functions are
$\Oscr(\epsilon^{2^r})$ at $r$--th normalization step).
In~\cite{Cel-Gio-Loc-2000}, such a reformulation of the procedure
constructing the Kolmogorov's normal form is shown to be highly
effective in practical applications; moreover, it is well suited to
locate invariant tori in Celestial Mechanics realistic problems
(see~\cite{Loc-Gio-2000}, \cite{Gab-Jor-Loc-2005},
\cite{Loc-Gio-2005}, \cite{Loc-Gio-2007} and~\cite{San-Loc-Gio-2011}).
Let us also stress that the Kolmogorov's normal form can be used so to
ensure the effective stability in a neighborhood of an invariant KAM
torus, because the drift motion of the eventual diffusion can be
estimated to be extremely slow (see~\cite{Mor-Gio-95}
and~\cite{Gio-Loc-San-2009}).

In~\cite{Stef-Loc-2012}, we have shown that equations of
type~(\ref{eq:pendolo-dissip-model-Newt}) can be treated in the more
general context of pseudo-Hamiltonian action--angle structures with
$n_1+n_2\ge 2$ degrees of freedom, where there are $n_2\ge 0$ fixed
additional frequencies and the friction terms are {\it linear and
  homogeneous} with respect to the $n_1\ge 1$ actions; therefore, by
using a technique of {\it quadratic} type, we proved the convergence
of the algorithm constructing the Kolmogorov's normal form adapted to
this pseudo-Hamiltonian framework, if the perturbation is small
enough. In the present work, we reformulate the constructive procedure
according to a {\it classical} formal scheme; moreover, we explicitly
calculate the expansions of the Hamiltonians defined up to a fixed
finite normalization step, by algebraic manipulations on a
computer. It is now rather common to say that such a method is
semi-analytic, where we mean that we are going to use a constructive
formal algorithm whose the convergence (at least for small
perturbations) might be ensured by an analytic rigorous proof, but we
limit us to show it, by directly checking the expansions produced on a
computer.

For the sake of completeness, let us recall that recently the
existence of quasi-periodic solutions for dissipative systems has been
proved also in the more general context of conformally symplectic
systems (see~\cite{Cal-Cel-dlL-2013.2}). Such a result is based on a
technique designed also to produce powerful applications to realistic
models. Furthermore, that approach can be extended so to describe also
the locally attracting dynamics in the neighborhood of the
quasi-periodic solutions, although the proof scheme does not ensure
the existence of any normal form (see~\cite{Cal-Cel-dlL-2013.3}).

\smallskip
This paper is organized as follows. In
section~\ref{sec:models}, we define the models, which will be studied
by our numerical explorations. In section~\ref{sec:method}, we adapt
the frequency map analysis method to dissipative systems and, as a
first stressing test, we compare our results about the breakdown
threshold of invariant tori for the dissipative standard map, with
those obtained by computing the Sobolev norms.
Section~\ref{sec:results-diss-forc-pend-model} is devoted to the
exploration of the dissipative forced pendulum model, by applying our
adaptation of the frequency analysis.  In section~\ref{sec:kolmog},
the algorithm constructing the Kolmogorov's normal form is adapted to
the general pseudo-Hamiltonian framework and it is applied to the
dissipative forced pendulum model; this is done to check the agreement
with some numerical results described in
section~\ref{sec:results-diss-forc-pend-model} and, also, to describe
some features of the local dynamics attracting to the invariant torus,
whose existence is ensured by the corresponding Kolmogorov's
normal form. Conclusions are drawn in section~\ref{sec:conclu}.

\section{Introducing the models: dissipative standard map and
forced pendulum}
\label{sec:models}

In the present work we consider two simple but fundamental systems:
the dissipative standard map and the forced pendulum with dissipation.

The standard map
$\Sscr_{\epsilon}:\reali\times\toro\mapsto\reali\times\toro$ is
certainly the most famous symplectic map; here, we add a dissipation
which is linear in the action variable. Thus, we consider the model
defined by the equations
\begin{equation}
  \left\{\begin{array}{rll}
    y'= & y + \epsilon \sin x - \eta(y-\Omega) & \quad y \in \reali \\
    x'= & x + y' & \quad {\rm mod}\ 2\pi
  \end{array}\right.
\ ,
\label{def:sm_dissipativa}
\end{equation}
where $\epsilon\ge 0$ is the perturbing parameter controlling the size
of the perturbation, $\eta \ge 0$ is the friction coefficient ruling
the dissipation rate and $\Omega$ is an external forcing frequency.
Let us remark that 
 when $\eta=0$ the
formula above covers also the usual definition of the conservative standard map
$\Sscr_{\epsilon}\,$.  Moreover, in the unperturbed case (i.e., when
$\epsilon=0$), the set $\{y=\Omega\,,\,x\in\toro\}$ is an invariant
global attractor of the dynamics and $\Omega$ is also the frequency
value of the angular motion on that torus.

The dissipative standard map has been widely studied, like for example
in~\cite{Ce-Fro-Le06}, where it is defined as follows:
\begin{equation}
  \left\{\begin{array}{rll}
    Y'= & \ds bY + c + \frac{\epsilon}{2\pi} \sin (2\pi X) &
          \quad Y \in \reali \\
    X'= & \ds X + Y' & \quad {\rm mod}\ 1
  \end{array}\right.
\ .
\label{def:sm_dissipativa_cell}
\end{equation}
In that case, the obvious correspondence between variables and parameters 
appearing
in~(\ref{def:sm_dissipativa}) and in~(\ref{def:sm_dissipativa_cell}) 
is given by the equations $x=2\pi X\,$,
$y=2\pi Y\,$, $\eta=1-b$ and $\eta\Omega=2\pi c\,$.
\smallskip

The second system considered here is the dissipative pseudo-Hamiltonian
model of the forced pendulum. In order to define it properly,
let us introduce the autonomous Hamiltonian describing the forced
pendulum (with the variable $q_2$ playing the role of time)
\begin{equation}
H_{\epsilon}(p_1,p_2,q_1,q_2)= \frac{{p_1}^2}{2}+ p_2 +
\epsilon\left[\cos q_1+\cos(q_1-q_2)\right]\ , 
\label{eq:Ham_pendolo}
\end{equation}
where $(p_1\,,\,p_2)\in{\reali}^2$, $(q_1\,,\,q_2)\in\toro^2$ and
$\epsilon$ is a small positive parameter. Let us simplify the
notation, by introducing the ``Hamiltonian vector field operator''
$\Hamvecfield\,$, which acts on a dynamical function
$g:\reali^n\times\toro^n\mapsto\reali$ (where $n$ is a generic number of
degrees of freedom) so that
\begin{equation}
\Hamvecfield(g)=
\left(-\frac{\partial g}{\partial q_1}\,,\,\ldots\,,\,
-\frac{\partial g}{\partial q_n}\,,\,\frac{\partial g}{\partial p_1}
\,,\,\ldots\,,\,\frac{\partial g}{\partial p_n}\right)
\ .
\label{eq:def-Hamvectorfield}
\end{equation}
Therefore, our pseudo-Hamiltonian model of the dissipative forced
pendulum is described by the following equation:
\begin{equation}
\left(\dot p_1,\dot p_2,\dot q_1,\dot q_2\right)=
\Hamvecfield\big(H_{\epsilon}\big)
-\eta\big(p_1-\Omega\,,\,0\,,\,0\,,\,0\big)\ ,
\label{eq:pendolo-dissip-pseudo-Ham}
\end{equation}
where the meaning of the symbols $\eta$ and $\Omega$ is the same as
in~(\ref{def:sm_dissipativa}).

The dissipative forced pendulum introduced above is substantially
defined by a system of three differential equations depending on the
variables~$q_1,\,q_2,\,p_1\,$; the evolution of the action~$p_2$ is
actually irrelevant, because it does not have any influence on the
behavior of the other variables. Moreover, once the law of motion
$t\mapsto\big(q_1(t),\,q_2(t),\,p_1(t)\big)$ is known, the function
$t\mapsto p_2(t)$ can be determined by computing an integral.  When
one is interested in investigating numerically the behavior induced by
the differential equation~(\ref{eq:pendolo-dissip-pseudo-Ham}), it is
natural to consider the corresponding Poincar\'e map.  This allows us
to reduce the numbers of variables from 3 to 2, by sampling the state
of the system at times which are multiple integers of the period of
the variable~$q_2\,$, that is~$2\pi$. In other words, we are going to
study the Poincar\'e map
$M_{\epsilon\,,\,\eta\,,\,\Omega}\,:\,\reali\times\toro\mapsto
\reali\times\toro\,$, that is defined so that
\begin{equation}
M_{\epsilon\,,\,\eta\,,\,\Omega}(p_1,q_1) =
\Phi_{\epsilon\,,\,\eta\,,\,\Omega}^{2\pi}(p_1,0,q_1)\ ,
\label{def:Poincare_map}
\end{equation}
where $\Phi_{\epsilon\,,\,\eta\,,\,\Omega}^{\delta}\,:\,
\reali\times\toro^2\mapsto\reali\times\toro^2$ is the $\delta$--time
flow induced by  equation~(\ref{eq:pendolo-dissip-pseudo-Ham}) and
we do not take into account its effect on $p_2\,$.

One can easily check that in the conservative case
$M_{\epsilon\,,\,0\,,\,\Omega}$ is a symplectic map. Let us recall
that, apart a further rescaling of the parameters, the dissipative
standard map is nothing but a very rough approximation of
$M_{\epsilon\,,\,\eta\,,\,\Omega}$ that is produced by a single step
of the so called semi-implicit Euler method, covering a time interval
equal to $2\pi\,$.

From a practical point of view, in all the numerical experiments
described in the present paper, the Poincar\'e map
$M_{\epsilon\,,\,\eta\,,\,\Omega}$ is approximated by a numerical
integration of the equations of
motion~(\ref{eq:pendolo-dissip-pseudo-Ham}), using the
Taylor\footnote{A software package implementing the numerical
  integration of the ordinary differential equations by means of the
  Taylor method is publicly available at the following website: {\tt
    http://www.maia.ub.es/{$\sim$}angel/soft.html}} method
(see~\cite{Jor-Zou-2005}). In our tests, such a software package is
able to numerically integrate the flow
$\Phi_{\epsilon\,,\,\eta\,,\,\Omega}^{2\pi}\,$, performing less than
$30$ steps; each step is affected by an uncertainity not greater
than the round-off error on {\tt double} type variables of the {\bf C}
programming language. The precision of the integration scheme could be
further improved, by using {\tt long double} type variables or
multiple precision arithmetic, that can be very well performed also by
using the {\tt TIDES} software package (see~\cite{Barrio-et-al-2012}
for an introduction). We consider that our numerical results should be
very slightly modified by such a further improvement and, so, it has
not been implemented.

An important feature of the dissipative systems is that they need a
relaxation time before converging to the invariant attractor. From the
computational point of view, this means that a certain number~$W$ of
preliminary iterations is necessary, in addition to those required by
the frequency map analysis.  In order to provide a criterion for the
choice of the value of $W\,$, let us consider the unperturbed case of
the dissipative standard map~(\ref{def:sm_dissipativa}) and assume the
initial value of the ordinate is $y_0\,$; then, one can easily check
that the sequence of the iterated points is such that
$y_n=(1-\eta)^n(y_0-\Omega)+\Omega\,$, $\forall\ n\in\naturali\,$.  In
a numerical experiment, the value of $y_0$ is determined so that the
initial point $(x_0,y_0)$ is rather close to the wanted invariant
attractor. Therefore, we define $W$ so that $(1-\eta)^n$ is at most of
the order of the machine precision $\forall\ n\ge W\,$, i.e.,
\begin{equation}
 W=\lceil -(15\log 10)/\log(1-\eta)\rceil\ ,
\label{e:wait_time}
\end{equation}
being $\lceil\alpha\rceil$ the smallest
integer greater than or equal to $\alpha\in\reali\,$. 

\section{Adapting the frequency map analysis to dissipative systems}
\label{sec:method}

\subsection{Frequency map analysis for Hamiltonian systems: a short
overview}
\label{sbs:ainf-Ham}

Since our investigation approach for dissipative systems is strongly
reminiscent of the method designed by Laskar to study conservative
systems (see, e.g.,~\cite{Laskar-99} and~\cite{Laskar-2005}), we think
that it is convenient to recall some of its features in the present
subsection. This will allow us to introduce our adaptation for
dissipative systems in a more natural way.

Let us consider an $n$--d.o.f. quasi-integrable system, described by
an analytic Hamiltonian
\begin{equation}
H(\vet{I},\vet{\theta})=h(\vet{I})+\epsilon f(\vet{I},\vet{\theta})\ ,
\label{eq:Ham-quasi-int}
\end{equation}
where $(\vet{I},\vet{\theta})\in\Gscr\times\toro^n$ (being
$\Gscr\subset\reali^n$ an open set) are action--angle variables.
According to KAM theory (see, e.g.,~\cite{Poeschel-1982}), if the
following conditions are satisfied:

(A)~the integrable part $h(\vet{I})$ is non-degenerate (i.e., the
determinant of the hessian of $h$ is different from zero
$\forall\ \vet{I}\in\Gscr$),

(B)~the parameter $\epsilon$ is small enough;

\noindent
then, there exists a diffeomorphism
$\Psi\,:\ \Bscr\times\toro^n\mapsto\Gscr\times\toro^n$ having the
following properties:

(I)~$\Psi(\vet{\omega},\vet{\phi})$ is invertible and it is
$\Cscr^{\infty}$ with respect to $\vet{\omega}\in\Bscr$ and analytic in
$\vet{\phi}\in\toro^n$,

(II)~there is a Cantor set $\Bscr_{\epsilon}\subset\Bscr$ such that
for each (diophantine) frequency $\vet{\omega}\in\Bscr_{\epsilon}$ the
law of motion
$t\mapsto(\vet{I}(t),\vet{\theta}(t))=\Psi(\vet{\omega},\vet{\omega} t)$ is
a solution of Hamilton's equations on an invariant (KAM) torus,

(III)~when $\Gscr$ is bounded, the Lebesgue measure of
$\Bscr\setminus\Bscr_{\epsilon}$ tends to zero for $\epsilon\to 0\,$.

\noindent
Let us recall that here the non-degeneracy condition on the integrable
part $h$ can be replaced by the so called isoenergetical
non-degeneracy~(see, e.g.,~\cite{Celletti-2010} for a
definition). Moreover, while practically doing numerical explorations,
the (very restrictive) smallness condition on the parameter $\epsilon$
can be ignored, because it is known that in a neighborhood of a
generic invariant KAM torus, there is a canonical transformation
leading the Hamiltomian to a form such that the above conditions~(A)
and~(B) are satisfied (see~\cite{Mor-Gio-95}).

Let $t\mapsto z(t)$ a signal (depending on time) in the complex plane,
where $z=z(\vet{I},\vet{\theta})$ is a function defined on the phase
space.  Let us suppose that the law of
motion~$t\mapsto(\vet{I}(t),\vet{\theta}(t))$ is quasi-periodic and is
characterized by the frequency vector $\vet{\omega}\,$, being its
corresponding orbit on an invariant KAM torus; then the property~(II)
above allows us to assume that the signal $t\mapsto
z(\vet{I}(t),\vet{\theta}(t))$ is as follows:
\begin{equation}
z(t)=\sum_{l=0}^{\infty}c_l\exp\big(\imunit\zeta_lt\big)\,,
\quad
{\rm where},\ \forall\ l\ge 0\,,\ \left\{
\begin{array}{l}
   c_l\in\complessi\\
   \exists\ \,\vet{k}_l\in\interi^{n}
   \ {\rm such\ that\ }\zeta_l=\vet{k}_l\cdot\vet{\omega}
  \end{array}\right.
\ .
\label{spettro-quasi-periodico}
\end{equation}
The basic software package implementing the numerical analysis of the
fundamental frequencies (see, e.g.,~\cite{LFC}) must allow us to
calculate a suitable truncation of the expansion above. Actually, the
values of the frequencies $\zeta_l$ are numerically found by looking
for the local maxima of the following function:
\begin{equation}
\sigma\mapsto \left|\frac{1}{2T}\int_{t_0-T}^{t_0+T}\diff
t\ z(t)\exp(-\imunit\sigma t)\,w\left(\frac{t-t_0}{T}\right)\right|\ ,
\label{def:integrale-an-in-freq}
\end{equation}
where $w$ is a {\sl weight function}, i.e., an analytic, non-negative,
even map $w:[-1,1]\mapsto\{0\}\cup\reali_+$ such that
$\int_{-1}^{1}w(u)\diff u=2\,$.  Moreover, $[t_0-T,t_0+T]$ is meant to
be a time interval where the signal $t\mapsto z(t)$ has been
preliminarly computed (usually, by a numerical integration
approximating the law of motion
$t\mapsto(\vet{I}(t),\vet{\theta}(t))\,$).  In practical applications,
it is natural to sample such a signal with $N$ uniform subintervals of
$[t_0-T,t_0+T]\,$, being their width equal to
$\Delta=2T/N\,$. Therefore, the integral appearing in
formula~(\ref{def:integrale-an-in-freq}) can be approximated by using
the trapezoidal rule or a similar quadrature formula.  Of course, one
expects that the numerical calculation of the Fourier
decomposition~(\ref{spettro-quasi-periodico}) becomes better and
better when the value of $T$ increases. We suggest to the reader (once
again) the reviews~\cite{Laskar-99} and~\cite{Laskar-2005} for the
careful discussion about the accuracy of the numerical results and their
dependence on the parameter $T\,$, $\Delta$ and the weight function
$w\,$. In all our numerical experiments (described below), we used the
Hanning window filter $w(u)=1+\cos(\pi u)\,$; here, we just recall that
with this kind of weight function, the difference between the computed
value of the frequency vector $\vet{\omega}$ and the true one is
$\Oscr(1/T^4)$ for $T\to\infty\,$.

A first test to check if an orbit lies on an invariant KAM torus can
be made by controlling that the Fourier
decomposition~(\ref{spettro-quasi-periodico}) of a corresponding
signal holds true, within the limitations due to the unavoidable
numerical errors. Let us remark that the frequency vector
$\vet{\omega}$ is not given {\it a priori}, but its detection is often
not so difficult, in practical applications.  For instance, let us
consider a quasi-integrable Hamiltonian of the
type~(\ref{eq:Ham-quasi-int}) and satisfying the conditions~(A)
and~(B); $\forall\ j=1,\,\ldots\,,\,n\,$, let us study the signal
$z(t)=I_j(t)\exp\big(\imunit\theta_j(t)\big)\,$; then the point
$\bar\sigma_T$ corresponding to the absolute maximum of the
function~(\ref{def:integrale-an-in-freq}) gives an approximation of
$\omega_j\,$, that gets more and more accurate for $T\to\infty\,$.

Another natural numerical investigation concerns  the local
regularity and invertibility of the {\it action-frequency map}
$\vet{I}\mapsto \vet{\omega}(\vet{I})$ such that
$\vet{\omega}(\vet{I})=\Psi^{-1}(\vet{I},\vet{\theta})$ for any fixed
value of $\vet{\theta}\in\toro^n\,$.  The frequency map analysis
mainly aims to obtain directly, in a numerical manner, the map
$\vet{I}\mapsto \vet{\omega}(\vet{I})\,$. The procedure, can be
summarized as follows: we first arbitrarily fix the initial values of
the angles $\vet{\theta}_0\,$; we pick up the initial actions
$\vet{I}_0$ from a regular grid $\Jscr$ of values. For each initial
condition, we consider the corresponding motion law
$t\mapsto(\vet{I}(t),\vet{\theta}(t))$ and we analyze the $n$ signals
$z_j(t)=I_j(t)\exp\big(\imunit\theta_j(t)\big)$ with
$t\in[t_0-T,t_0+T]$, $\forall\ j=1,\,\ldots\,,\,n\,$; for each signal,
we find the value of $\omega_j$ corresponding to the absolute maximum of
the function~(\ref{def:integrale-an-in-freq}). Following this
procedure, we can then calculate the frequency
$\vet{\omega}=\vet{\omega}(\vet{I_0})$ for all the initial values of the
actions $\vet{I}_0\in\Jscr\,$.

The analysis of the frequency map can distinguish among three
different dynamical situations:

(a)~when the values of the initial conditions
$(\vet{I}_0,\vet{\theta}_0)$ are such that the corresponding motions
are chaotic $\forall\ \vet{I}_0\in\Jscr\,$, then the map
$\vet{I}_0\mapsto \vet{\omega}(\vet{I}_0)$ looks highly irregular;

(b)~when the initial conditions are such that the orbits are on the
regular manifolds inside a resonant region (these are the so called
``librational'' maximal tori in the neighborhood of a stable
equilibrium point or an elliptic lower dimensional torus), some
components $\omega_j$ of the frequency vector are constant while
$\vet{I}_0$ is changed;

(c) when the initial conditions are in a region (nearly) filled by KAM
tori, a thin enough enlargement of the frequency map highlights a
quasi-linear and invertible behavior. This is in agreement with the
property~(I) (that is described above and proved
in~\cite{Poeschel-1982}) joined with the approach described in,
e.g.,~\cite{Mor-Gio-95}: the signature of the existence of a KAM torus
is the local regularity and invertibility of the {\it
  action--frequency map} $\vet{I}\mapsto \vet{\omega}(\vet{I})$ such that
$\vet{\omega}(\vet{I})=\Psi^{-1}(\vet{I},\vet{\theta})$ for any fixed
value of $\vet{\theta}\in\toro^n\,$.

\begin{figure}
\begin{center}
\includegraphics[width=0.95\textwidth]{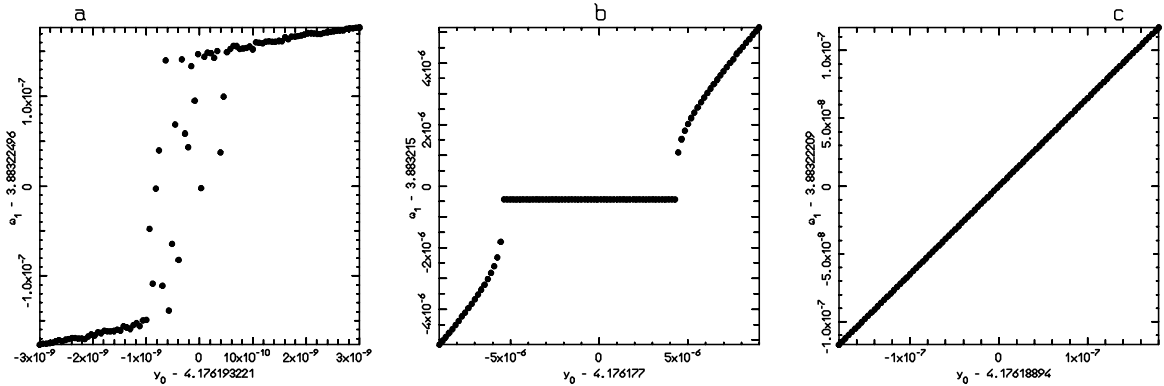}
\end{center}
\caption{Frequency map analysis of the standard map
  $\Sscr_{\epsilon}\,$, with $\epsilon=0.97\,$. In the plots above,
  three different sets of orbits are considered; each orbit is
  computed starting from an initial condition of the type
  $(x,y)=(0,y_0)$, with $y_0$  in abscissa. The
  corresponding value of the main frequency $\omega_1$ has been
  calculated by analyzing the signal $n\mapsto y_n\exp(\imunit x_n)$
  with $0\le n\le N\,$ (see the text for further definitions and
  details). For the experiments in
  Figure~\ref{fig:dis_mappa_freq}a we fixed $N=2^{18}$, while for
  Figures~\ref{fig:dis_mappa_freq}b--c
  $N=2^{16}\,$. Figure~\ref{fig:dis_mappa_freq}a refers to some orbits
  in a neighborhood of the chaotic zone related to the resonance
  $610/987\,$; in the central part of
  Figure~\ref{fig:dis_mappa_freq}b, some chains of regular islands
  surrounding the stable periodic orbit of frequency $2\pi\,377/610$
  are considered; Figure~\ref{fig:dis_mappa_freq}c focuses on a
  neighborhood of the ``golden'' invariant torus of frequency
  $\omega_1=2\pi(\sqrt{5}-1)/2\,$.}
\label{fig:dis_mappa_freq}
\end{figure}

These three different regimes can be sharply highlighted with some
numerical experiments on symplectic maps. For instance, let us
consider the standard map $\Sscr_{\epsilon}\,$, as it is defined by
formula~(\ref{def:sm_dissipativa}) when $\eta=0\,$. The results
plotted in Figure~\ref{fig:dis_mappa_freq} are obtained by analyzing the
signal $n\mapsto z(n)$ with $z(n)=y_n\exp(\imunit x_n)\,$, where the
pair $(y_n,x_n)$ is obtained by $n$ iterations of
$\Sscr_{\epsilon}\,$, starting from the initial condition
$(y_0,x_0)\,$. Let us remark that we are assuming that the signal is
sampled in a trivial way, so that the ``elapsed time'' between an
iteration of the standard map and the next one is $\Delta=1\,$. By
looking at the definitions in~(\ref{spettro-quasi-periodico})
and~(\ref{def:integrale-an-in-freq}), one can easily realize that
changing the definition of $\Delta$ would imply a harmless rescaling
of the found value of the frequency $\vet\omega$ by a factor $1/\Delta\,$.
We can appreciate that the archetypical behaviors described at the
points (a)--(c) are clearly detected by the numerical experiments,
whose  results are plotted in Figures~\ref{fig:dis_mappa_freq}a--c,
respectively.

\subsection{Frequency map analysis for dissipative systems}
\label{sbs:ainf-diss}

In some dissipative systems, there is just one global attractor for the
dynamics. For instance, if we consider the unperturbed
dissipative forced pendulum, described by the
equation~(\ref{eq:pendolo-dissip-pseudo-Ham}) setting $\epsilon =
0$ in~(\ref{eq:Ham_pendolo}), the solution for the motion of the
action $p_1$ can be written as
\begin{equation}
p_1(t) = \big(p_1(0)-\Omega\big)\exp(-\eta t) + \Omega\ .
\label{eq:sol-in-azione-rotatore-dissipativo}
\end{equation}
By the way, let us remark that the Hamiltonian~(\ref{eq:Ham_pendolo})
when $\epsilon = 0$ describes nothing but a rotator plus a
clock. Looking at the equation above, it is obvious that
$p_1(t)\to\Omega$ for $t\to\infty\,$. The motion law on the global
attractor is given by the following equations:
\begin{equation}
p_1(t) = \Omega\ ,
\qquad
q_1(t) = \Omega t+q_1(0)\ .
\qquad
q_2(t) = t\ .
\label{eq:sol-su-attrattore-rotatore-dissipativo}
\end{equation}
In the perturbed case (i.e., when $\epsilon\neq 0$), one can provide
examples of weakly dissipative systems, where there are more than one
single attractor (see, e.g.,~\cite{Cel-Chi-2008}). Since each basin of
attraction usually contains open sets of the phase space, one
immediately realizes that the study of the map
$\vet{\omega}(\vet{I})=\Psi^{-1}(\vet{I},\vet{\theta})$ loses sense
for dissipative systems. In fact, there are many initial values of the
actions $\vet{I}$ corresponding to the same final frequency
$\vet{\omega}\,$, therefore, the action--frequency map is obviously
not invertible even when the final attractor is an invariant torus.

However, the trivial example of the unperturbed case can help us to
explain the simple idea underlying our new approach. The
solution~(\ref{eq:sol-su-attrattore-rotatore-dissipativo}) highlights
that the frequency of the invariant attractor is $\omega_1=\dot
q_1=\Omega\,$, thus, when $\epsilon=0$ the map
$\Omega\mapsto\omega_1(\Omega)$ is obviously regular and invertible,
because it is the identity.  It is natural to expect that
such a map somehow remains regular and invertible also in the perturbed case
for small values
of~$\epsilon$. Actually, this is guaranteed by the main result of Celletti
and Chierchia in~\cite{Cel-Chi-2009}, at least for systems of type of
the dissipative forced pendulum, that is defined by the
equations~(\ref{eq:Ham_pendolo})--(\ref{eq:pendolo-dissip-pseudo-Ham}). In
fact, when the perturbing terms are small enough, Theorem~1
of~\cite{Cel-Chi-2009} claims also the following relation between the
frequency $\omega_1$ of the quasi-periodic motion (on an invariant
torus) and the external forcing frequency $\Omega\,$:
\begin{equation}
\Omega=\omega_1\left(1+\Oscr(\epsilon^2)\right)\ ,
\label{eq:rel-frequenze-moto-forzante-esterna}
\end{equation}
when $\vet{\omega}\in\Dscr_{\gamma,\tau}\,$, being $\Dscr_{\gamma,\tau}$ the
set of diophantine numbers such that
\begin{equation}
\big|n_1\omega_1+n_2 \omega_2| 
\ge \frac{\gamma}{\big|n_1\big|^\tau}
\qquad
\forall\ (n_1,n_2)\in\interi^2,\ n_1\neq 0
\ ,
\label{def:num_diofantei}
\end{equation}
for some fixed values $0<\gamma<1$ and $\tau\ge 1\,$. Moreover, the
function $\omega_1\mapsto\Omega(\omega_1)$ is Whitney\footnote{Let us
  recall that a function $g:A\subset\reali^{m_1}\mapsto\reali^{m_2}$
  is said to be Whitney $\Cscr^{k}$, if it is a restriction on $A$ of
  a $\Cscr^{k}(\reali^{m_1})$ function.} $\Cscr^{\infty}$ on the
Cantor set $\Dscr_{\gamma,\tau}\,$. Therefore, the
equation~(\ref{eq:rel-frequenze-moto-forzante-esterna}) leads us to
conclude that the map $\Omega\to\omega_1(\Omega)$ is regular and
locally invertible in the neighborhood of an invariant torus, if
$\epsilon$ is small enough. As it has been claimed in remark~(v) of
section~(1.1) of~\cite{Cel-Chi-2009}, it is expected that such kind of
results can be extended to systems with more degrees of freedom. Thus,
we {\it conjecture} that when a dissipative system is governed by the
following equations of motion
\begin{equation}
\big(\dot{\vet{I}},\dot{\vet{\theta}}\big)=
\Hamvecfield\big(H\big)
-\eta\big(\vet{I}-\vet{\Omega}\,,\,0\big)\ ,
\label{eq:sistema-pseudo-Ham}
\end{equation}
and its Hamiltonian part $H$ satisfies the hypotheses~(A)--(B)
(described in subsection~\ref{sbs:ainf-Ham}), then there
exists a diffeomorphism $\Xi(\vet{\omega},\vet{\phi})$ 
which satisfies the same properties~(I)--(III) (holding for the diffeomorfism
$\Psi$~of the conservative case), with the action vector $\vet{I}$
replaced by the external forcing frequency vector $\vet{\Omega}\,$.

The previous discussion leads us to conclude that the frequency map
analysis can be adapted to the dissipative systems of the
type~(\ref{eq:sistema-pseudo-Ham}), by simply using the external
forcing frequency vector $\vet{\Omega}$ instead of the initial value
of the action $\vet{I}_0\,$. Actually, here we can consider a set of
motions, each of them corresponds to a different value of
$\vet{\Omega}\,$, got from a regular grid $\Ascr\,$. For the sake of
simplicity, we postpone to the next
sections further details about the procedure
calculating the corresponding frequencies of the motion on the
attractors. By analogy with the points (a)--(c) of 
subsection~\ref{sbs:ainf-Ham}, we {\it guess that also here the analysis
  of the frequency map can distinguish among three different dynamical
  situations}:

($a^{\prime}$)~when the values of the external forcing frequency
vector are such that (for some set of initial conditions) the
corresponding orbits converge to a {\it strange attractor}
$\forall\ \vet{\Omega}\in\Ascr\,$, then the map $\vet{\Omega}\mapsto
\vet{\omega}(\vet{\Omega})$, should look highly irregular; actually,
both a strange attractor and a chaotic orbit have fractal dimension
larger than the degrees of freedom (see Figure~5 in~\cite{CDLS10}),
therefore, it is natural to argue that the behavior will be the same
as for the chaotic motions in Hamiltonian systems (see point~(a) of
the previous subsection~\ref{sbs:ainf-Ham});

($b^{\prime}$)~when the external forcing frequency vectors are such
that some orbits converge to attractors that are regular manifolds
inside a resonant region, some components $\omega_j$ of the frequency
vector are constant as $\vet{\Omega}$ varies; for instance,
this statement is well supported by the numerical experiments on the
dissipative standard map; in that case, let us recall that each
periodic orbit of fixed frequency $\omega_1$ exists if and only if the
external frequency parameter
$\Omega\in[\Omega_{\omega_1;\,-},\Omega_{\omega_1;\,+}]\,$; moreover, when
the order of resonance related to $\omega_1$  increases, the interval
$[\Omega_{\omega_1;\,-},\Omega_{\omega_1;\,+}]$ gets smaller and smaller
(see, e.g.,~\cite{Cel-Diru-2011});

($c^{\prime}$) when the values of the external forcing frequency
vector are in a region of the regular grid $\Ascr$ such that the
corresponding attractors are invariant KAM tori, then a thin enough
enlargement of the frequency map should highlight a quasi-linear and
invertible behavior. This is in agreement with our conjecture about
the systems governed by the equation of
motion~(\ref{eq:sistema-pseudo-Ham}), when its Hamiltonian part $H$
satisfies the hypotheses~(A)--(B) (described in the previous
subsection~\ref{sbs:ainf-Ham}).

Finally, let us remark that it is natural to expect that the
behaviors described at the previous
points~($a^{\prime}$)--($c^{\prime}$) should hold, also when the
non-degeneracy condition~(A) of subsection~\ref{sbs:ainf-Ham} is
replaced by the weaker one we considered in~\cite{Stef-Loc-2012}
(i.e., condition~(b) of theorem~3.1). Moreover, it
is natural to guess that the analysis of the frequency map should
highlight the same situations also for dissipative maps, that are
obtained as a Poincar\'e map of the continuous flow induced by
equations of motion of the type~(\ref{eq:sistema-pseudo-Ham}).

\subsection{Numerical experiments on the dissipative standard map}
\label{sbs:ainf-diss-std-map}

The interpretation of the frequency map described in the previous
subsection allows us to investigate the breakdown threshold of
invariant tori for dissipative maps. This allows us to submit our
method to some challenging test, because we can compare our results
with some existing ones in literature.

\begin{figure}
\begin{center}
\includegraphics[width=145mm]{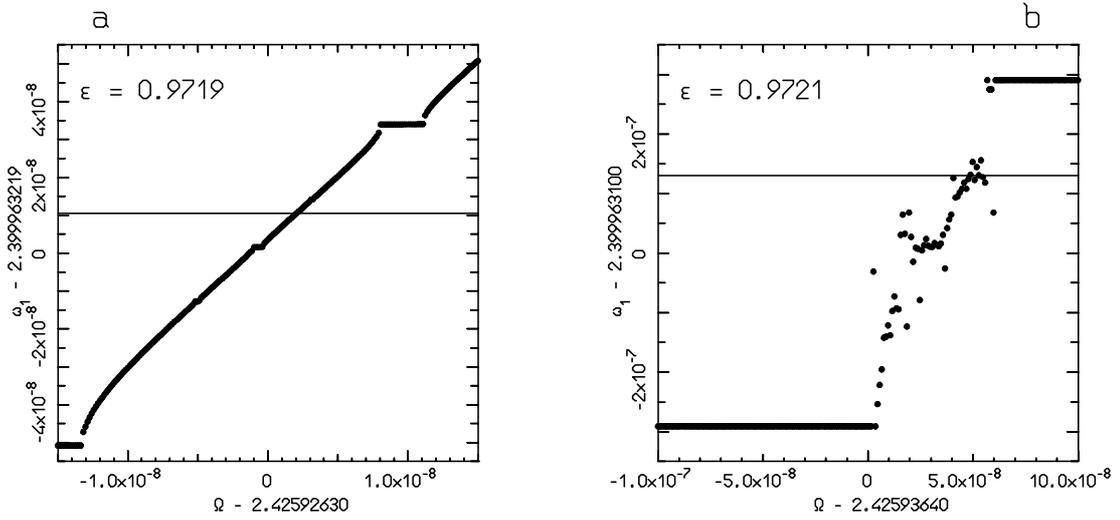}
\end{center}
\caption{Frequency analysis for the dissipative standard
  map~(\ref{def:sm_dissipativa}), with the friction parameter
  $\eta=0.1\,$. The range of abscissas (related to the external
  frequency values of $\Omega)$ has been determined so to focus
  on a neighborood of the golden  value
  $\omega_1/2\pi=2-\golden=(3-\sqrt{5})/2\,$.  The left plot refers to
  the case~$\epsilon=0.9719$ and the right one to $\epsilon=0.9721\,$. }
\label{fig:sm_break_eta0.1}
\end{figure}

Let us focus on the dissipative standard
map~(\ref{def:sm_dissipativa}). Figure~\ref{fig:sm_break_eta0.1} shows
the frequency maps $\Omega\mapsto\omega_1(\Omega)$ for
$\epsilon=0.9719$ and $\epsilon=0.9721$; in both cases the friction
parameter~$\eta$ has been set equal to $0.1\,$.  Those maps have been
drawn by analyzing signals of the type $n\mapsto z(n)$ with
$z(n)=y_n\exp(\imunit x_n)\,$, where the pair $(y_n,x_n)$ is obtained
by $n$ iterations of the dissipative standard
map~(\ref{def:sm_dissipativa}), starting from the initial condition
$(y_0,x_0)=(\Omega,0)\,$. Each point plotted in
Figure~\ref{fig:sm_break_eta0.1} is actually related to a single
analysis of an orbit corresponding to the value of $\Omega$ reported
in abscissa; such an analysis considers all the values of index $n$
ranging in $[W,N+W]$, with $W$ given by formula~(\ref{e:wait_time})
and $N=2^{19}=524288\,$. This means that we perform a ``waiting''
number $W$ of preliminary iterations (that are needed to let the orbit
approach very closely an invariant attractor), before starting the
calculation of the value $\omega_1(\Omega)$ corresponding to the
absolute maximum of the map~(\ref{def:integrale-an-in-freq}). 
Let us recall also that the integral appearing in
formula~(\ref{def:integrale-an-in-freq}) is approximated numerically
by using the trapezoidal rule with $N$ subintervals (all with the same
width) in $[t_0-T,t_0+T]\,$, being $t_0=W+N/2$ and $T=N/2\,$;
moreover, the ``weight'' function $w$ is the Hanning window filter
$w(u)=1+\cos(\pi u)\,$.

Before discussing the results we need some preliminary remarks about
the definition of the frequencies. First, let us recall that there is
a special class among the Diophantine frequencies, that is given by
``noble'' numbers, having their continued fraction expansion ending
with only~$1$; in particular we will consider here the ``golden
number''~$\golden=[1;1^{\infty}]=(\sqrt 5 + 1)/2\simeq 1.618\,$.
Moreover, in the case of the dissipative standard map, as for any
discrete time map, frequencies that differ by an integer multiple
of~$2\pi$ are equivalent, in the sense that their dynamics are
undistinguishable. In the special case of the dissipative standard
map, it is also obvious that the dynamics is also $2\pi$-periodic in
the action. These properties mean that tori whose frequencies differ
by any multiple of~$2\pi$ are equivalent, and in particular all the
``golden tori'' with frequencies~$2\pi\golden\,$, $2\pi(\golden-1)\,$,
$2\pi(\golden-2)\,$, $\ldots$ have the same shape and break in the
same way.  Moreover, for the dissipative standard map, the dynamics is
also invariant when changing $x \rightarrow -x$, $y \rightarrow -y$
and $\Omega \rightarrow -\Omega$; this also implies that tori having
opposite frequencies~$\omega_1$ and~$-\omega_1$ also behave in the
same way. We thus decided to perform our numerical experiments on the
torus with frequency~$\omega_1/2\pi=2-\golden=(3-\sqrt{5})/2\simeq
0.381966\,$, which is the only golden torus with positive
frequency~$\omega_1$ in $[0,\pi]\,$. In
Figure~\ref{fig:sm_break_eta0.1} (both on the left and on the right),
the thin horizontal lines correspond to the value of the ``golden
torus'' frequency $\omega_1=2\pi(2-\golden)=2\pi[(3-\sqrt{5})/2]\,$.

According to our discussion in
subsection~\ref{sbs:ainf-diss}, we are led to conclude that the
attractor related to the golden mean frequency exists, if the map
$\Omega\mapsto\omega_1(\Omega)$ looks regular (i.e., quasi-linear) in a
small neighborhood of the intersection with the thin horizontal line;
otherwise, when the map shows sudden jumps where it is crossing the
thin line, then that invariant torus does not exist. The left panel
of Figure~\ref{fig:sm_break_eta0.1} clearly shows that ``golden torus''
still persists for $\epsilon=0.9719\,$, while the right panel makes
evident that it is destroyed when the parameter ruling the
perturbation is $\epsilon=0.9721\,$. This allows us to conclude that the
breakdown threshold $\epsilon_c$ should be in the interval
$(0.9719\,,\,0.9721)$ when $\omega_1=2\pi(2-\golden)$ and $\eta=0.1\,$.  By
the way, let us remark that in Figure~\ref{fig:sm_break_eta0.1} the
large ``plateaus'' appearing in the left plot and in the right one
correspond to the resonant values of $\omega_1/(2\pi)$ equal to
$2584/6765\,$, $1597/4181\,$ (left), $987/2584\,$, and $1597/4181\,$
(right).

\begin{table}
\begin{center}
\def\arraystretch{1.6}
\begin{tabular}{|c||c|c|c|} \hline
$\omega_1/2\pi$ & $\eta=0.1$ & $\eta=0.2$ & $\eta=0.5$ \\ \hline
$2-\golden$ 
& $\epsilon_c=$ 0.972 $\pm$ $10^{-4}$  & $\epsilon_c=$ 0.973 $\pm$ $10^{-3}$
& $\epsilon_c=$ 0.979 $\pm$ $10^{-3}$  \\
\hline
$\left[0;2,5,3,1^{\infty}\right]$
& $\epsilon_c=$ 0.846 $\pm$ $10^{-3}$ & $\epsilon_c=$ 0.859 $\pm$ $10^{-3}$
& $\epsilon_c=$ 0.918 $\pm$ $10^{-3}$  \\
\hline
\end{tabular}
\end{center}
\vspace{3mm}
\caption{Study of the breakdown threshold for a couple of invariant
  tori for the dissipative standard
  map~(\ref{def:sm_dissipativa}). The critical values~$\epsilon_{c}$ (of
  the small parameter $\epsilon$) have been computed by using our approach
  based on frequency analysis. Such critical values of the breakdown
  threshold are obtained for different values of the
  dissipation coefficient~$\eta$ and for the invariant tori of
  frequencies $2-\golden$ and
  $\left[0;2,5,3,1^{\infty}\right]\,$.}
\label{tab:sm_tab_eps_eta}
\end{table}

In Table~\ref{tab:sm_tab_eps_eta} we collect some results obtained by
applying our method to compute the critical values $\epsilon_c$ of the
breakdown threshold for a pair of invariant tori and a few different
values of the friction parameter $\eta\,$. Let us stress that we
repeatedly use the same procedure described for both the cases of the
golden torus with $\omega_1/(2\pi)=2-\golden$ and of the frequency
$\omega_1/(2\pi)=\left[0;2,5,3,1^{\infty}\right]\simeq 0.4567\,$. The
cases studied here can be directly compared with those considered by
Calleja and Celletti (see Tables~I--III in~\cite{Cal-Cel-2010}).
Since the results listed in Table~\ref{tab:sm_tab_eps_eta} are in
agreement with both those based on the computation of the Sobolev
norms and those obtained by applying the Greene's method, we consider
that this comparison strongly support the validity of our approach,
that was heuristically motivated in the previous
subsection~\ref{sbs:ainf-diss}.

A more detailed comparison of the results with those provided
in~\cite{Cal-Cel-2010} highlights that the most performing method (to
determine the breakdown threshold) is that based on the computation of
the Sobolev norms; in fact, it provides the largest number of
significant digits (about five). We emphasize that our approach can be
nicely visualized (as in Figure~\ref{fig:sm_break_eta0.1}), but, as an
evident drawback, it is not easy to make the whole procedure very
automatic. For instance, the determination of a suitable range of
abscissas often requires many trials and errors; moreover, the numbers
of trials significantly increases when a high precision is
required. This is because we limited ourselves to compute the
breakdown threshold up to the third significant digit in all the cases
listed in Table~\ref{tab:sm_tab_eps_eta}, except for that illustrated in
Figure~\ref{fig:sm_break_eta0.1}.

\section{Numerical results about the dissipative forced pendulum}
\label{sec:results-diss-forc-pend-model}

Let us now focus on the equation of
motion~(\ref{eq:pendolo-dissip-pseudo-Ham}) for the dissipative forced
pendulum, where the Hamiltonian part $H_{\epsilon}$ is defined
in~(\ref{eq:Ham_pendolo}). The aim of this section is also to
determine the values of some parameters (for instance, the breakdown
threshold $\epsilon_c$ and the external forcing frequency $\Omega$),
which must be known in advance, before starting any explicit
calculation of Kolmogorov's normalization algorithm, that will be
discussed in the next section.

\subsection{Breakdown of invariant tori in the dissipative forced pendulum}

It is natural to adopt exactly the same approach used in
subsection~\ref{sbs:ainf-diss-std-map} to study the dissipative
standard map, in order to investigate the behavior of the Poincar\'e
map~(\ref{def:Poincare_map}), related to the dissipative forced
pendulum.  Namely, after having somehow fixed the values of the
parameters $\epsilon\,$, $\eta$ and $\Omega\,$, we can produce
frequency maps by analyzing the signals of the type $n\mapsto z(n)$
with $z(n)=y_n\exp(\imunit x_n)\,$, where the pair $(y_n,x_n)$ is
obtained by $n$ iterations of the dissipative map
$M_{\epsilon\,,\,\eta\,,\,\Omega}\,$, defined
in~(\ref{def:Poincare_map}). Each signal $n\mapsto z(n)$ is analyzed
so to determine the value $\omega_1$ of the absolute maximum point of
the map~(\ref{def:integrale-an-in-freq}); such an integral is
numerically approximated in the same way as we did in
subsection~\ref{sbs:ainf-diss-std-map}.  In particular, the endpoints
of the interval $[t_0-T,t_0+T]$ are fixed so that $t_0=2\pi(W+N/2)$
and $T=2\pi(N/2)\,$, where $W$~is given by formula~(\ref{e:wait_time})
and $N=2^{16}=65536\,$.

\begin{figure}
\begin{center}
\includegraphics[width=145mm]{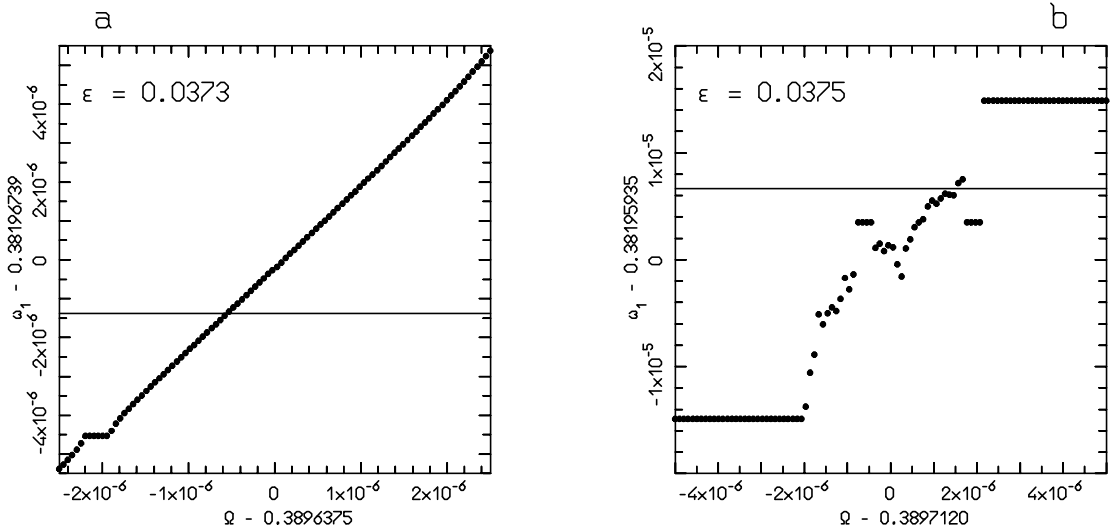}
\end{center}
\caption{{Frequency analysis for the dissipative forced pendulum,
    defined by the
    equations~(\ref{eq:Ham_pendolo})--(\ref{eq:pendolo-dissip-pseudo-Ham}),
    with the friction parameter $\eta=0.1\,$. The range of abscissas
    (related to the external frequency values of $\Omega)$ has been
    determined so to focus on a neighbourood of
    $\omega_1=2-\golden=(3-\sqrt{5})/2\,$, being $\golden$ the ``golden
    mean'' value. The left plot refers to the case~$\epsilon=0.0373$ and
    the right one to $\epsilon=0.0375\,$.}}
\label{fig:breakdown-torodoro-eta0.1}
\end{figure}

As a first numerical investigation about the dynamics of the
dissipative forced pendulum, we study the breakdown of the invariant
``golden torus''. Let us remark that since we decided to sample the
continuous dynamics with a timestep~$\Delta=2\pi\,$, the
relation~(\ref{eq:rel-frequenze-moto-forzante-esterna}) still holds,
but with frequencies~$\omega_1$ now in the interval~$[-0.5,0.5]$.  The
frequency maps $\Omega\mapsto\omega_1(\Omega)$ for $\epsilon=0.0373$
and $\epsilon=0.0375$ are plotted in
Figure~\ref{fig:breakdown-torodoro-eta0.1}, in both cases the friction
parameter $\eta$ is equal to $0.1\,$. Both in the left plot and in the
right one, the thin horizontal lines correspond
to the value of the ordinate equal to the frequency
$\omega_1=2-\golden=(3-\sqrt{5})/2 \simeq 0.381966\,$. According to
the discussions in
subsections~\ref{sbs:ainf-diss}--\ref{sbs:ainf-diss-std-map}, we can
provide a clear interpretation of the results illustrated in
Figure~\ref{fig:breakdown-torodoro-eta0.1}: the breakdown threshold of
the wanted invariant torus is $\epsilon_{c}=0.0374\pm 10^{-4}$.

In the case of the dissipative forced pendulum, we think that it is
interesting to study the dependence of the breakdown threshold
$\epsilon_{c}(\omega_1,\eta)$ on the friction coefficient $\eta\,$.
More precisely, we want check if the function
$\eta\mapsto\epsilon_{c}(\omega_1,\eta)$ behaves according to the
KAM--like analytical estimates. 
For completeness, we recall below the discussion about some functional
properties of the theoretical threshold $\epsilon^{\star}$ (which
depends on many parameters characterizing the system)  included in
sect.~3 of~\cite{Stef-Loc-2012}.

\begin{enumerate}
\renewcommand{\itemsep}{0pt}
\item[(A)] There is a range of ``small'' values of the friction
  parameter, with $0\le\eta\le\eta_1^{\star}\,$, for which
  $\epsilon^{\star}$ is a constant.
\item[(B)] There is an ``intermediate'' range of values of $\eta\,$,
  with $\eta_1^{\star}\le\eta\le\eta_2^{\star}\,$, for which the
  function $\eta\mapsto\epsilon^{\star}$ is increasing.
\item[(C)] The value of $\eta_2^{\star}$ depends on both the
  non-resonance assumptions about the frequency vector and those
  guaranteeing the non-degeneracy; actually, $\eta_2^{\star}\le
  10^5\,\eta_1^{\star}$ and the limit case $\eta_2^{\star}=
  10^5\,\eta_1^{\star}\,$ should hold true just when the latter
  conditions are much weaker than the former ones.
\item[(D)] If $\eta\ge\eta_2^{\star}\,$, the function
  $\eta\mapsto\epsilon^{\star}$ is decreasing; in particular, when the
  value of the friction parameter is very large, then
  $\epsilon^{\star}=\Oscr(1/|\eta|)$ for $|\eta|\to\infty\,$.
\end{enumerate}

\noindent Let us recall that the previous points (A)--(D) cover also
the so called anti-dissipative case with $\eta<0\,$, because all the
analytical estimates depend just on the absolute value of the friction
parameter.

\begin{figure}
\begin{center}
\def\arraystretch{1.2}
\begin{tabular}{|c|l|}\hline
$\eta$  &  $\phantom{0.0}\epsilon_{c}$ \\ \hline
0.005 & 0.02780 \\
0.01  & 0.02807 \\
0.02  & 0.0294 \\
0.05  & 0.0322 \\
0.1   & 0.0374 \\
0.2   & 0.0497 \\
0.5   & 0.1020 \\
1.0   & 0.240 \\ \hline
\end{tabular}
\hspace{0.5cm}
\parbox[c]{9cm}{\includegraphics[width=8cm]{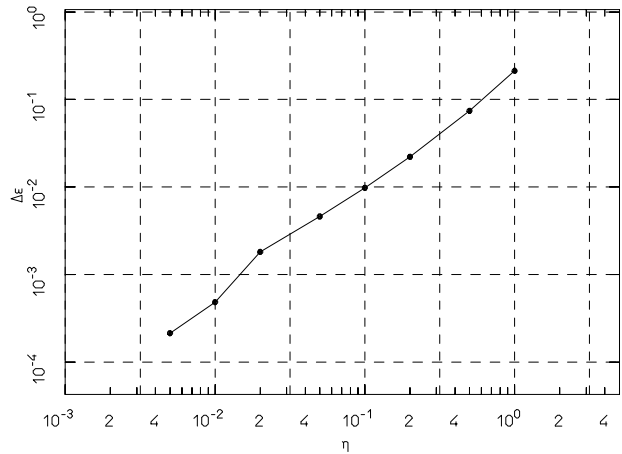}}
\end{center}
\caption{Dissipative forced pendulum: variation of the breakdown
  threshold~$\epsilon_{c}$ as a function of the dissipation
  rate~$\eta$ for the golden torus of frequency $\omega_1=2-\golden=
  (3-\sqrt 5)/2\,$. The plot is in log--log scale and shows in
  ordinate the values of
  $\Delta\epsilon=\epsilon_c(2-\golden,\eta)-{\bar\epsilon}_c\,$, where
  ${\bar\epsilon}_c\simeq 0.0275856$ is the breakdown threshold of the
  conservative case.}
\label{fig:tab-graf-eta-epsilon-pend-torodoro} 
\end{figure}

Here, we limit ourselves to investigate the function
$\eta\mapsto\epsilon_{c}(\omega_1,\eta)$ in the case of the ``golden
torus'' with frequency $\omega_1=2-\golden=(3-\sqrt{5})/2$ (adopted
for consistency with the experiment on the dissipative standard map),
so to compare the predictions for the theoretical breakdown threshold
$\epsilon^{\star}$ with the behavior of the numerical one
$\epsilon_{c}\,$.  Figure~\ref{fig:tab-graf-eta-epsilon-pend-torodoro}
includes (on the left) a table with some values of the correspondence
$\eta\mapsto\epsilon_{c}\,$, numerically determined by applying our
frequency analysis approach to the study of the dissipative forced
pendulum. When the friction coefficient $\eta$ gets smaller and
smaller then the corresponding value of $\epsilon_c$ seems to converge
to ${\bar\epsilon}_c\simeq 0.0275856\,$, that is the breakdown
threshold of the golden torus in the conservative case (see,
e.g.,~\cite{Gov-Cha-Jau-97}). Since the map $\eta\mapsto\epsilon_{c}$
looks regular and we can guess that it is an even function (let us
recall that the value of $\epsilon^{\star}$ is preserved by the
simmetry $\eta\mapsto-\eta\,$), the fact that the scaling law for
$\eta\to 0$ is clearly superlinear suggests that
$\eta\mapsto\epsilon_{c}(2-\golden,\eta)$ has a quadratic minimum in
the origin. This is in agreement with the behavior of the theoretical
breakdown threshold $\epsilon^{\star}$ described at points~(A)--(B).
Moreover, the plot in logarithmic scale (on both axes) of
Figure~\ref{fig:tab-graf-eta-epsilon-pend-torodoro} highlights the
(approximately) linear growth of the numerical breakdown threshold
$\epsilon_{c}$ when $\eta\in[0.02\,,\,1]\,$.

 It would be very interesting to study more widely the function
 $\eta\mapsto\epsilon_{c}(2-\golden,\eta)\,$, by investigating a set
 of values of the dissipative parameter~$\eta$ larger than that
 considered in Figure~\ref{fig:tab-graf-eta-epsilon-pend-torodoro}.
 Unfortunately, a further extension of a few orders of magnitude for
 the plotted values of $\eta$ is very demanding from a computational
 point of view, because of two different reasons. For small values of
 $\eta\,$, the calculation of the ordinate
 $\Delta\epsilon=\epsilon_c(2-\golden,\eta)-{\bar\epsilon}_c$ is
 meaningful just when the breakdown threshold
 $\epsilon_c(2-\golden,\eta)$ is determined with many significant
 digits and this is a hard task for our method, as discussed in
 subsection~\ref{sbs:ainf-diss-std-map}. On the other hand, when a
 large value of the friction coefficient is considered, a plot similar
 to those reported in Figure~\ref{fig:breakdown-torodoro-eta0.1} can
 require a too long CPU--time. Indeed, a few experiments with our
 numerical integrator allowed us to check that its internal time-step
 $\Delta t$ is automatically set so that $\Delta t=\Oscr(1/\eta)$ for
 $\eta\to\infty\,$.  Thus, we are far from being able to detect the
 behavior described at points~(C)--(D).  Let us remark that numerical
 experiments on the dissipative standard map are not affected by such
 a computational limitation (that is induced by any refined method
 integrating numerically the flow of the dissipative forced pendulum),
 when the value of $\eta$ is increased. Thus, the same exploration
 could be done for mappings, but, as far as we know, this case is not
 yet covered by a theorem providing a careful description of the
 behavior of the breakdown threshold as a function of the friction
 coefficient, in a similar way to what is reported at points (A)--(D).

\subsection{Numerical determination of the forcing frequency}
\label{sbs:appr-forc-freq}

In order to perform explicitly the algorithm constructing the
Kolmogorov's normal form (as described in the next section) for
dissipative systems, we need to preliminarly determine the external
forcing frequency. To fix the ideas, we limit ourselves to consider
again the dynamics of the dissipative forced pendulum.  The aim of
this subsection is to determine, for a fixed invariant torus, the
corresponding value of the parameter~$\Omega$ (appearing in the
equation~(\ref{eq:pendolo-dissip-pseudo-Ham}), where the Hamiltonian
part $H_{\epsilon}$ is given in~(\ref{eq:Ham_pendolo})).  As discussed
above, the frequency map provides the frequency $\omega_1$ of the
quasi-periodic motion on an invariant torus as a function of the
parameter~$\Omega\,$. Now, the problem is the following: we fix the
frequency~$\omega_1=\omega_1^{*}$ related to an invariant torus and we
need to approximate numerically the corresponding forcing
frequency~$\Omega^{*}$.  Thus, denoting again the frequency map by
$\Omega\mapsto\omega_1(\Omega)\,$, we want to find numerically the
solution~$\Omega^{*}$ of the equation
$$
\omega_1(\Omega^{*})=\omega_1^{*}\ .
$$
This requires to invert the frequency map, or, more simply, to find
the real zero of the function
$$
f(\Omega)=\omega_1(\Omega)-\omega_1^{*}\ .
$$
For this purpose, we implement explicitly a Newton's method which,
as it is well known, is an iterative method to find numerically the
solutions for this kind of problems.  Let us stress that we expect to
find a locally unique solution $\Omega^{*}$ of the equation
$\omega_1(\Omega)-\omega_1^{*}=0\,$, because we obviously apply
Newton's method for values of the parameter $\epsilon$ smaller than
the breakdown threshold $\epsilon_{c}$ of the invariant torus related
to the frequency $\omega_1^{*}$. Therefore, in a neighborhood of the
unknown value $\Omega^*$, the function $\omega_1(\Omega)$ has a
quasi-linear behavior, which looks strictly monotone, except in the
resonant zones (see the left plots in
Figure~\ref{fig:sm_break_eta0.1}--\ref{fig:breakdown-torodoro-eta0.1}
and the discussions about them). Thus, if the initial approximation
belongs to the region about the solution where the map
$\Omega\mapsto\omega_1(\Omega)$ looks mostly quasi-linear and
monotone, Newton's method is expected to be very efficient.

In order to be more definite, in the following we describe our procedure 
in detail. We denote with~$\tilde\Omega^{(j)}$,
the $j$-th approximation of the solution~$\Omega^{*}$; then, the
single step of  Newton's algorithm, applied
to~$f(\Omega)=\omega_1(\Omega)-\omega_1^{*}$ prescribes that the next
approximation~$\tilde\Omega^{(j+1)}$ is given by
\begin{equation}
\tilde\Omega^{(j+1)}=\tilde\Omega^{(j)}
            - \frac{\omega_1\big(\tilde\Omega^{(j)}\big)-\omega_1^{*}}
              {\omega_1\big((1+\alpha)\tilde\Omega^{(j)}\big)-
                \omega_1\big(\tilde\Omega^{(j)}\big)}\,
              \alpha\tilde\Omega^{(j)}\ ,
\label{eq:Newton-step}
\end{equation}
where the derivative $f^{\prime}\big(\tilde\Omega^{(j)}\big)$ is
replaced by the finite difference
$\big[\omega_1\big((1+\alpha)\tilde\Omega^{(j)}\big)
  -\omega_1\big(\tilde\Omega^{(j)}\big)\big]
/\big(\alpha\tilde\Omega^{(j)}\big)$ and $\alpha$ is a small parameter
to be conveniently fixed so to ensure the numerical stability of this
procedure. Of course, in formula~(\ref{eq:Newton-step}) the values of
$\omega_1\big(\tilde\Omega^{(j)}\big)$ and
$\omega_1\big((1+\alpha)\tilde\Omega^{(j)}\big)$ are numerically
calculated, by using the frequency analysis. We stop the iterations
when the relative correction on the value of $\Omega$ is below a fixed
precision~$\beta\,$, i.e., when
\begin{equation}
\frac{\left|\tilde\Omega^{(j+1)}-\tilde\Omega^{(j)}\right|}
{\left|\tilde\Omega^{(j+1)}\right|+\left|\tilde\Omega^{(j)}\right|}< \beta\ ,
\label{diseq:stop-Newton}
\end{equation}
where $\beta$ is a small parameter that can be conveniently chosen so
to be not much greater than the machine precision (let us recall that
this is about $2.2\times 10^{-16}$ for the standard {\tt double}
precision type numbers).

Let us remark that it is very unlikely that at some step the finite
difference $\big[\omega_1\big((1+\alpha)\tilde\Omega^{(j)}\big)
  -\omega_1\big(\tilde\Omega^{(j)}\big)\big]
  /\big(\alpha\tilde\Omega^{(j)}\big)$ be close to $0\,$, because it
occurs that both $(1+\alpha)\tilde\Omega^{(j)}$ and
$\tilde\Omega^{(j)}$ are in the same resonant region. In fact, the
sizes of the resonant ``plateaus'' are smaller and smaller, when
approaching the invariant torus; this fact can be seen in the left
plots of
Figures~\ref{fig:sm_break_eta0.1}--\ref{fig:breakdown-torodoro-eta0.1}
and it has been clearly shown in the conservative framework (see,
e.g., the numerical investigations in~\cite{LF96}). Thus, we limited
ourselves to include a test in our code, to stop the running if the
finite difference above is too small. This event is so rare that it
never happened in our calculations; from a practical point of view, in
such a case of failure, one has to look for a better initial
approximation $\tilde\Omega^{(0)}$ before restarting the procedure.

For instance, let us discuss an explicit case. We want to determine
the value of the external forcing frequency such that the equations of
motion~(\ref{eq:pendolo-dissip-pseudo-Ham}) has the golden torus as an
invariant attractor, when the values of the parameters are fixed so that
$$
\epsilon=0.03\ ,
\qquad
\eta=0.1\ .
$$
Let us recall that the value of the small parameter $\epsilon$ is
chosen smaller than the breakdown threshold related to
$\omega_1^*=2-\golden=(3-\sqrt{5})/2$ and $\eta=0.1$ (see the 
corresponding value of $\epsilon_c$ in the table appearing in
Figure~\ref{fig:tab-graf-eta-epsilon-pend-torodoro}). 
We fix $\alpha=10^{-6}$ and $\beta=10^{-15}$ and we start the
Newton's algorithm taking $\tilde\Omega^{(0)}=\omega_1^*$ as initial
approximation.  In this case, the
algorithm ends successfully after just $5$ steps with 
\begin{equation}
\tilde\Omega^{(5)}=0.3870821721708347\ .
\label{eq:val-Omegone-appr-5}
\end{equation}

As an internal test of our result, we performed the decomposition of
the Fourier spectrum as in
formula~(\ref{spettro-quasi-periodico}). Actually, we considered the
motion on the invariant attractor for the dissipative forced pendulum
defined by equation~(\ref{eq:pendolo-dissip-pseudo-Ham}), with
$\epsilon=0.03\,$, $\eta=0.1$ and $\Omega=\Omega^*\simeq\tilde\Omega^{(5)}$,
with the value of $\tilde\Omega^{(5)}$ given in
(\ref{eq:val-Omegone-appr-5}). Since the golden torus is expected to
be the invariant attractor, we tried to express every frequency
$\zeta_l$ as a linear combination of the components of the vector
$\vet\omega=\big((3-\sqrt{5})/2\,,\,1\big)\,$. The relevant quantities
involved in the decomposition of the Fourier spectrum are listed in
Table~6.2 of~\cite{Stefanelli-2011}, where the numerical
results definitely show that the invariant attractor is the golden
torus related to the frequency
$\omega_1^*=2-\golden=(3-\sqrt{5})/2\,$.

\section{Semi-analytic approach constructing the normal form for invariant tori}\label{sec:kolmog} 

In order to describe the procedure constructing explicitly the
Kolmogorov's normal form related to an invariant quasi-periodic
attractor, it is convenient to reformulate the pseudo-Hamiltonian
model of the dissipative forced pendulum (defined by the
equations~(\ref{eq:Ham_pendolo})--(\ref{eq:pendolo-dissip-pseudo-Ham}))
in a suitably more general context. For this purpose, let us introduce
three non-negative integer numbers $n_1\,$, $n_2$ and $K\,$; among
them, both $n_1$ and $K$ are strictly positive. Let $n=n_1+n_2$ be the
number of degrees of freedom of the system described by the following
equations of motion:
\begin{equation}
\left(\dot{\vet{p}},\dot{\vet{q}}\right)=
\Hamvecfield\left(H^{(0)}\right)-
\eta\big(\vet{p}-\vet{\Omega}^{(0)}\,,\,\vet{0}\big)\ ,
\label{eq:vec-field-H0}
\end{equation}
where~$\vet{\Omega}^{(0)}\in\reali^n$ is the external forcing frequency vector
and the Hamiltonian part
\begin{equation}
H^{(0)}(\vet{p},\vet{q})=\vet{\omega}\cdot\vet{p}+
\sum_{l=0}^{\infty}\sum_{s=0}^{\infty}f_l^{(0,s)}(\vet{p},\vet{q})\ .
\label{eq:def-H0}
\end{equation}
In the equation above, as usual, $\vet{\omega}\in\reali^n$ must be
regarded as a fixed frequency vector, while
$f_{l}^{(0,s)}\in\Pgot_{l,sK}$ $\forall$ $l\ge 0$ and $s\ge 0\,$,
where we denote $\Pgot_{l,sK}$ the class of functions which are homogeneous
polynomials of degree $l$ in $p_{1}\,$, $\ldots\,$, $p_{n_1}\,$, do
not depend on the $n_2$ actions $p_{n_1+1}\,$, $\ldots\,$, $p_{n}$ and
are trigonometric polynomials of degree $sK$ with respect to the
angles $\vet{q}\in\toro^n\,$. Let us remark that also the Hamiltonian 
of the forced pendulum $H_{\epsilon}$ defined in~(\ref{eq:Ham_pendolo})
can be expressed in the form~(\ref{eq:def-H0}) with the following
values of the integer parameters: $n_1=1\,$, $n_2=1$ and $K=2\,$. In
fact, after having performed a translation\footnote{Let us recall that
  $(p_1,p_2,q_1,q_2)\mapsto(p_1+\omega_1,p_2,q_1,q_2)$ is a canonical
  transformation. Moreover, we emphasize that in this special case the
  definition of $H^{(0)}(\vet{p},\vet{q})$ here is given by avoiding
  the introduction of a new symbol instead of $p_1\,$, by abuse of
  notation.} of the action coordinate $p_1$ so that
$p_1=p_1+\omega_1\,$, it is enough to put
$\vet{\omega}=(\omega_1,1)\,$,
$\vet{\Omega}^{(0)}=\vet{\Omega}-(\omega_1,0)=(\Omega-\omega_1,0)\,$,
$f_{0}^{(0,1)}=\epsilon[\cos q_1+\cos(q_1-q_2)]\,$,
$f_{2}^{(0,0)}=\frac{1}{2}p_1^2$ and $f_{l}^{(0,s)}=0$
$\,\,\forall\ (l,s)\neq (0,1),\,(2,0)\,$.

Our goal is to determine an accurate approximation of a canonical
transformation $\psi^{(\infty)}$, such that in the new coordinates
$(\vet{P},\vet{Q})=\big(\psi^{(\infty)}\big)^{-1}(\vet{p},\vet{q})$ the
equations of motion~(\ref{eq:vec-field-H0}) are transformed to the
following form:
\begin{equation}
\big(\dot{\vet{P}},\dot{\vet{Q}}\big)=
\Hamvecfield\left(H^{(\infty)}\right)-
\eta\big(\vet{P}\,,\,\vet{0}\big)\ ,
\label{eq:vec-field-Hinfty}
\end{equation}
where the new Hamiltonian $H^{(\infty)}$ is in Kolmogorov's normal
form, i.e.,
\begin{equation}
H^{(\infty)}(\vet{P},\vet{Q})=\vet{\omega}\cdot\vet{P}+
\sum_{l=2}^{\infty}\sum_{s=0}^{\infty}f_{l}^{(\infty,s)}(\vet{P},\vet{Q})\ ,
\label{eq:def-Hinfty}
\end{equation}
with $f_{l}^{(\infty,s)}\in\Pscr_{l,sK}$ $\forall\ l\ge 2\,$. In
words, the Kolmogorov's normal form is such that its part depending on
the angles is at least quadratic with respect to the actions.
Therefore, the invariance of the torus
$\{\vet{P}=\vet{0}\,,\,\vet{Q}\in\toro^n\}$ immediately follows from the
equations~(\ref{eq:vec-field-Hinfty})--(\ref{eq:def-Hinfty}). Of
course, that invariant torus is densely filled by a quasi-periodic
orbit characterized by the frequencies vector $\vet{\omega}\,$.

\subsection{Adapting the standard Kolmogorov's normalization formal algorithm to dissipative equations with friction terms that are linear and homogeneous with respect to the actions}\label{sbs:Kolm_norm_form}

Let us describe the generic $r$--th step of the adapted Kolmogorov's
normalization algorithm. We start from equations of motion of
type
\begin{equation}
\left(\dot{\vet{p}},\dot{\vet{q}}\right)=
\Hamvecfield\left(H^{(r-1)}\right)-
\eta\big(\vet{p}-\vet{\Omega}^{(r-1)}\,,\,\vet{0}\big)\ ,
\label{eq:vec-field-Hr-1}
\end{equation}
where the Hamiltonian part can be expanded as follows:
\begin{equation}
H^{(r-1)}(\vet{p},\vet{q})=\vet{\omega}\cdot\vet{p}+
\sum_{l=0}^{\infty}\sum_{s=0}^{\infty}f_l^{(r-1,s)}(\vet{p},\vet{q})\ ,
\label{eq:def-Hr-1}
\end{equation}
with $f_l^{(r-1,s)}\in\Pgot_{l,sK}$ $\forall$ $l\ge 0$ and $s\ge
0\,$. Moreover, we require that the Taylor--Fourier series above is
``well ordered''. This assumption is not restrictive, actually we mean
that, $\forall$ $l\ge 0$ and $s\ge 0\,$, each term
$c_{\vet{j},\vet{k}}^{(r-1)}\,\vet{p}^{\vet{j}}\,\exp(i\vet{k}\cdot\vet{q})$
appearing in the expansion of $f_l^{(r-1,s)}$ is such that
$|\vet{j}|=l$ and $(s-1)K<|\vet{k}|\le sK\,$, where we used the common
multi-index notation $\vet{p}^{\vet{j}} = p_1^{j_1}\cdot \ldots \cdot
p_{n_1}^{j_{n_1}}$ and $|\cdot|$ is the $l_1$--norm, for instance,
$|\vet{k}|=|k_1|+\ldots +|k_n|\,$.

The $r$--th normalization step is split in two separate steps.  We
first remove part of the unwanted terms via a canonical transformation
having $\chi_1^{(r)}(\vet{q})=X^{(r)}(\vet{q})+
{\vet{\csi}}^{(r)}\cdot\vet{q}$ as generating function. Lemma~2.4
of~\cite{Stef-Loc-2012} ensures us that, after performing such a first
canonical transformation, the new equations of motion have the
following form:
\begin{equation}
\big(\dot{\vet{p}},\dot{\vet{q}}\big)=
\Hamvecfield\big(\hat H^{(r)}\big)-
\eta\big(\vet{p}-\hat{\vet{\Omega}}^{(r)}\,,\,\vet{0}\big)\ ,
\label{eq:vec-field-hatHr}
\end{equation}
where
\begin{equation}
\hat{\vet{\Omega}}^{(r)}=\vet{\Omega}^{(r-1)}+\vet{\xi^{(r)}}
\label{eq:per-hatomegone}
\end{equation}
and the new Hamiltonian part is given by
\begin{equation}
\hat H^{(r)}=\exp\Big(\Lie_{\chi_1^{(r)}}\Big)H^{(r-1)}-\eta X^{(r)}\ .
\label{eq:def-funzionale-hatHr}
\end{equation}
In order to avoid the proliferation of too many symbols, starting from
equation~(\ref{eq:vec-field-hatHr}) we do not introduce another set of
variables for the new coordinates introduced after each canonical
transformation; this is done by abuse of notation. Moreover, in the
functional equation~(\ref{eq:def-funzionale-hatHr}) the Lie series
operator $\exp\big(\Lie_{\chi_1^{(r)}}\big)$ appears, where
$\Lie_{\chi} g=\poisson{g}{\chi}\,$ and $\poisson{\cdot}{\cdot}$ is
the classical Poisson bracket, $g$ a generic function defined on the
phase space and $\chi$ any generating function.

Since we point to a Hamiltonian part of type~(\ref{eq:def-Hinfty}),
first, we determine the generating function $\chi_1^{(r)}$ so to
remove both the main perturbing terms of degree $0$ and those that are
linear with respect to the actions but do not depend on the
angles. Thus, we solve with respect to $X^{(r)}(q)$ and $\csi^{(r)}$
the equations
\begin{equation}
\sum_{i=1}^{n}\omega_i\frac{\partial\,X^{(r)}}{\partial q_i}(\vet{q})
-\eta X^{(r)}(\vet{q})+
\sum_{s=1}^r f^{(r-1,s)}_0(\vet{q})=0
\ ,
\qquad
\Cscr^{(r-1)}\vet{\csi}^{(r)}\cdot\vet{p}+f^{(r-1,0)}_1(\vet{p})=0\ ,
\label{eq:chi_1}
\end{equation}
where the $n_1\times n_1$ matrix $\Cscr^{(r-1)}$ is such that
$\frac{1}{2}\,(p_1,\,\ldots\,,\,p_{n_1}) \cdot
\Cscr^{(r-1)}(p_1,\,\ldots\,,\,p_{n_1})^{T} = f^{(r-1,0)}_2\,$ (let us
recall that both $f^{(r-1,0)}_1$ and $f^{(r-1,0)}_2$ do not depend on
the angles, because $f^{(r-1,0)}_1\in\Pgot_{1,0}$ and
$f^{(r-1,0)}_2\in\Pgot_{2,0}\,$).  After having expanded $\sum_{s=1}^r
f^{(r-1,s)}_0$ in Fourier series as
$$
\sum_{s=1}^r f^{(r-1,s)}_0(\vet{q})=\sum_{{\vet{k} \in \interi^n}\atop{0<|\vet{k}|\le rK}}
c_{\vet{0},\vet{k}}^{(r-1)}\exp(i\vet{k}\cdot\vet{q})\ ,
$$
we can easily write the solution of the first homological equation
appearing in~(\ref{eq:chi_1}), i.e.,
\begin{equation}
X^{(r)}(\vet{q})=\sum_{{\vet{k} \in \interi^n}\atop{0<|\vet{k}|\le rK}}
\frac{c_{\vet{0},\vet{k}}^{(r-1)}}{i \vet{k}\cdot\vet{\omega} + \eta}
        \exp(i \vet{k}\cdot\vet{q})\ .
\label{eq:sol-omol-per-X}
\end{equation}
Let us emphasize that the solution above is well defined when the
friction coefficient $\eta\neq 0\,$; in the conservative case (i.e.,
$\eta=0$), it is enough to use the non-resonance
condition~(\ref{diseq:diofantea-finoa-rK}), that will be explicitly
adopted to solve the second homological equation. Moreover,
one can easily realize that the second equation in~(\ref{eq:chi_1}), 
defines a linear system in the $n_1$ unknowns
$(\xi_1,\,\ldots\,,\,\xi_{n_1})\,$, because
$f^{(r-1,0)}_1\in\Pgot_{1,0}$ and $f^{(r-1,0)}_2\in\Pgot_{2,0}\,$;
this linear system can always be solved, provided that
\begin{equation}
\det\left(\Cscr^{(r-1)}\right)\neq 0\ .
\label{diseq:nondeg-cond-passo-r}
\end{equation}
Of course, the definition of $\vet{\xi}^{(r)}\in\reali^n$ is completed by
setting $\xi_{n_1+1}=0,\,\ldots\,,\,\xi_{n}=0\,$.

We must now provide the expressions of the functions
$\hat f_l^{(r,s)}$ appearing in the expansion of the new Hamiltonian
part
\begin{equation}
\hat H^{(r)}(\vet{p},\vet{q})=
\vet{\omega}\cdot\vet{p}+
\sum_{s\ge 0}\sum_{l\ge 0}\hat f_l^{(r,s)}(\vet{p},\vet{q})
\ ,
\label{eq:hatH^r}
\end{equation}
where $\hat H^{(r)}$ is defined by the functional
equation~(\ref{eq:def-funzionale-hatHr}). To this aim, we will
redefine many times the same quantity without changing the symbol. In
our opinion, such a repeated abuse of notation has two advantages:
first, this makes easier to understand the final calculation of $\hat
f_l^{(r,s)}$ instead of using one single very complicated formula;
second, the description of the algorithm is more similar to its
translation in a programming code. For instance, mimicking the {\bf C}
language, with the notation $a\pluseq b$ we mean that the previously
defined quantity $a$ is redefined as $a=a+b\,$.  Therefore, we
initially define
\begin{equation}
\hat f_l^{(r,s)}=f_l^{(r-1,s)}
\qquad\ \forall\ l\ge 0\ {\rm and}\ s\ge 0\ .
\label{eq:hatf_l^rs_def_1}
\end{equation}
To take into account the Poisson bracket of the generating function
with $\vet{\omega}\cdot\vet{p}$ and the contribution of the term
$-\eta X^{(r)}$, we put
\begin{equation}
\hat f_0^{(r,0)}\pluseq \vet{\omega}\cdot\vet{\csi^{(r)}}\ ,
\qquad
\hat f_0^{(r,s)}=0
\quad\ \forall\ 1\le s\le r\ .
\label{eq:hatf_l^rs_def_2}
\end{equation}
Then, we consider the contribution of the terms generated by the Lie
series applied to each function $f_l^{(r-1,s)}$ as follows:
\begin{equation}
\hat f_{l-j}^{(r,s+jr)}\pluseq
\frac{1}{j!}\Lie_{\chi_1^{(r)}}^jf_l^{(r-1,s)}
\qquad\ \forall\ l\ge 1\,,\ s\ge 0\ {\rm and}\ 1\le j\le l\ .
\label{eq:hatf_l^rs_def_3}
\end{equation}
Looking at
formul\ae~(\ref{eq:hatf_l^rs_def_1})--(\ref{eq:hatf_l^rs_def_3}), one
can easily check that $\hat f_l^{(r,s)}\in\Pgot_{l,sK}$ $\forall$
$l\ge 0$ and $s\ge 0\,$. We perform now a {\it ``reordering of the
  terms''}, by moving the monomials in the expansions of $\hat
f_l^{(r,s)}$ to each others, in such a way that, at the end, each term
of type ${\hat
  c}_{\vet{j},\vet{k}}^{(r)}\,\vet{p}^{\vet{j}}\,\exp(i\vet{k}\cdot\vet{q})$
belonging to the Taylor--Fourier (finite) series of the so redefined
functions $\hat f_l^{(r,s)}$ has degree $|\vet{j}|=l$ in the actions
and a trigonometric degree $|\vet{k}|\in\big((s-1)K\,,\,sK\big]\,$;
thus, it still holds true that $\hat f_l^{(r,s)}\in\Pgot_{l,sK}\,$.

In the second half of the $r$--th step of the adapted Kolmogorov's
normalization algorithm, by using another canonical transformation, we
remove the part of the perturbation up to the order of magnitude $r$
that actually depends on the angles and it is linear in the actions.
For this purpose, we are going to determine a generating function
$\chi_2^{(r)}$ that is linear with respect to the actions; therefore,
lemma~2.3 of~\cite{Stef-Loc-2012} ensures us that, after performing
the canonical transformation related to $\chi_2^{(r)}$, the new
equations of motion have the following form:
\begin{equation}
\big(\dot{\vet{p}},\dot{\vet{q}}\big)=
\Hamvecfield\big(H^{(r)}\big)-
\eta\big(\vet{p}-\vet{\Omega}^{(r)}\,,\,\vet{0}\big)\ ,
\label{eq:vec-field-Hr}
\end{equation}
where
\begin{equation}
\vet{\Omega}^{(r)}=\hat{\vet{\Omega}}^{(r)}
\label{eq:per-omegone}
\end{equation}
and the new Hamiltonian part is given by
\begin{equation}
H^{(r)}=\exp\Big(\Lie_{\chi_2^{(r)}}\Big)
\big(\hat H^{(r)}-\eta\,\hat{\vet{\Omega}}^{(r)}\cdot\vet{q}\big)
+\eta\,\hat{\vet{\Omega}}^{(r)}\cdot\vet{q}\ .
\label{eq:def-funzionale-Hr}
\end{equation}

In order to approach a Hamiltonian part of type~(\ref{eq:def-Hinfty}),
it is convenient to solve the following equation with respect to
$\chi_2^{(r)}(p,q)\,$:
\begin{equation}
\sum_{i=1}^{n}\omega_i
\frac{\partial\,\chi_2^{(r)}}{\partial q_i}(\vet{p},\vet{q})+
\sum_{s=1}^r \hat f^{(r,s)}_1(\vet{p},\vet{q})=0 \ .
\label{eq:chi_2}
\end{equation}
After having expanded $\sum_{s=1}^r \hat f^{(r,s)}_1$ in Fourier
series so that
\begin{equation}
\sum_{s=1}^r \hat f^{(r,s)}_1(\vet{p},\vet{q})=
\sum_{{\vet{j} \in \naturali^{n_1}}\atop{|\vet{j}|=1}}\,
\sum_{{\vet{k} \in \interi^n}\atop{0<|\vet{k}|\le rK}}
\hat c_{\vet{j},\vet{k}}^{(r)}\,\vet{p}^{\vet{j}}\,\exp(i\vet{k}\cdot\vet{q})\ ,
\end{equation}
if the following non-resonance condition is satisfied:
\begin{equation}
|\vet{k}\cdot\vet{\omega}|>0
\qquad
\forall\ \vet{k}\in\interi^n\ {\rm with}\ 0<|\vet{k}|\le rK\ ,
\label{diseq:diofantea-finoa-rK}
\end{equation}
then we can easily write the solution of the second homological
equation~(\ref{eq:chi_2}), i.e.,
\begin{equation}
\chi_2^{(r)}(\vet{p},\vet{q})=
\sum_{{\vet{j} \in \naturali^{n_1}}\atop{|\vet{j}|=1}}\,
\sum_{{\vet{k} \in \interi^n}\atop{0<|\vet{k}|\le rK}}
\frac{\hat c_{\vet{j},\vet{k}}^{(r)}}{i \vet{k}\cdot\vet{\omega}}
\,\vet{p}^{\vet{j}}\,\exp(i \vet{k}\cdot\vet{q})\ .
\label{eq:sol-omol-per-chi2}
\end{equation}
Similarly to what we have done previously, we now provide
the expressions of the functions $f_l^{(r,s)}$ appearing in the
expansion of the new Hamiltonian part:
\begin{equation}
H^{(r)}(\vet{p},\vet{q})=
\vet{\omega}\cdot\vet{p}+\sum_{s\ge 0}\sum_{l\ge 0} f_l^{(r,s)}(\vet{p},\vet{q})
\ ,
\label{eq:H^r}
\end{equation}
where $H^{(r)}$ is defined in~(\ref{eq:def-funzionale-Hr}). We
initially define
\begin{equation}
f_l^{(r,s)}=\hat f_l^{(r,s)}
\qquad\ \forall\ l\ge 0\ {\rm and}\ s\ge 0\ .
\label{eq:f_l^rs_def_0}
\end{equation}
The terms due to the expression
$\exp\big(\Lie_{\chi_2^{(r)}}\big)(-\eta\,\hat{\vet{\Omega}}^{(r)}\cdot\vet{q})
+\eta\,\hat{\vet{\Omega}}^{(r)}\cdot\vet{q}$ do not depend on the actions
and they contribute to the new Hamiltonian part as follows:
\begin{equation}
f_0^{(r,jr)}\pluseq
\frac{1}{j!}L_{\chi_2^{(r)}}^{j-1}\left(\eta\,\sum_{i=1}^{n_1}
{\hat\Omega}_i^{(r)}\frac{\partial\,\chi_2^{(r)}}{\partial p_i}\right)
\quad\ \forall\ j\ge 1\ .
\label{eq:f_l^rs_def_1}
\end{equation}
In view of the terms generated by the Lie series applied to
$\vet{\omega}\cdot\vet{p}$ and the second homological
equation~(\ref{eq:chi_2}), we put
\begin{equation}
f_1^{(r,s)}=0
\quad\ \forall\ 1\le s\le r\ ,
\qquad
f_1^{(r,jr)}\pluseq
\frac{1}{j!}L_{\chi_2^{(r)}}^{j-1}
\left(\sum_{i=1}^n\omega_i\frac{\partial\,\chi_2^{(r)}}{\partial q_i}\right)
\quad\ \forall\ j\ge 2\ .
\label{eq:f_l^rs_def_2}
\end{equation}
Moreover, the contribution of the Lie series applied to the rest of the
Hamiltonian $\hat H^{(r)}$ implies that
\begin{equation}
f_{l}^{(r,s+jr)}\pluseq
\frac{1}{j!}L_{\chi_2^{(r)}}^j\hat f_l^{(r,s)}
\quad\ \forall\ l\ge 0\,,\ s\ge 0\ {\rm and}\ j\ge 1\ .
\label{eq:f_l^rs_def_3}
\end{equation}
Finally, we perform a new {\it ``reordering of the terms''}, so that
at the end the functions $f_l^{(r,s)}\in\Pgot_{l,sK}$ (appearing in
expansion~(\ref{eq:H^r})) contain just monomials of type
$c_{\vet{j},\vet{k}}\,\vet{p}^{\vet{j}}\,\exp(i\vet{k}\cdot\vet{q})$
with degree $|\vet{j}|=l$ in the actions and trigonometric degree
$|\vet{k}|\in\big((s-1)K\,,\,sK\big]\,$.

Let us recall that the canonical transformation $\Kscr^{(r)}$
inducing the Kolmogorov's normalization up to the step $r$ is
explicitly given by
\begin{equation}
\Kscr^{(r)}(\vet{p},\vet{q})=\exp L_{\chi_2^{(r)}}\bigg(\exp L_{\chi_1^{(r)}}
\Big(\ldots\exp L_{\chi_2^{(2)}}\big(\exp L_{\chi_1^{(2)}}
\>(\vet{p},\vet{q})\big)\ldots\Big)\bigg).
\label{eq:trasf_Kolmogorov}
\end{equation}
This conclude the $r$--th step of the algorithm that can be further
iterated. Let us stress that the next step can be completely carried
out, if both the non-degeneracy
condition~(\ref{diseq:nondeg-cond-passo-r}) and the non-resonant
inequality~(\ref{diseq:diofantea-finoa-rK}) still hold true, when the
index $r$ is replaced with $r+1\,$. Actually, for what concerns the
former assumption, this is usually ensured by requiring that both the
quadratic part of the initial Hamiltonian is non-degenereate (i.e.,
$\det(\Cscr^{(0)})\neq 0$) and the parameter $\epsilon$ is small
enough. Moreover, the latter non-resonance condition is usually
satisfied for all indexes $r\,$, provided the chosen frequency vector
$\vet{\omega}$ is Diophantine.

Let us emphasize that terms having different orders of magnitude with
respect to the small parameter $\epsilon$ are not handled separately  in
our expansions. In particular, we prescribed to perform operations
like the ``reordering of the terms'', which explicitly requires to sum
contributions corresponding to the same polynomial degree and Fourier
harmonic, but with different orders in $\epsilon\,$.  The main
advantage of this formulation is to save most of the memory
occupation, when the algorithm is translated in any programming
language (see the discussion at the end of section~4.1
of~\cite{Gab-Jor-Loc-2005}). As a consequence of this gain in
memory handling, more normalization steps can be performed (and in a
faster way); let us stress that this can definitely improve the final
accuracy of the results. From a practical point of view, when
$H_{\epsilon}$ is expressed in the initial form~(\ref{eq:def-H0}), the
parameter $\epsilon$ must be replaced by its numerical
value. Therefore, in all the expansions written in the present
subsection (whenever they are converging in some suitable domains),
the sup-norm of the functions of type $f_{l}^{(r,s)}$ and $\hat
f_{l}^{(r,s)}$ is geometrically decreasing with respect to both the
polynomial degree $l$ and the index $s\,$, that is related to the
their trigonometric degree $sK\,$.

\subsection{Semi-analytic results}
\label{sbs:semi-analytic-results}

\subsubsection{Checking the explicit construction of the Kolmogorov's normal form}\label{sss:check-Kolm-norm-form}

It could be astonishing that both the equations of motion
(\ref{eq:vec-field-Hr}) and the Hamiltonian part~(\ref{eq:H^r})
defined at the $r$-th normalization step have exactly the same
structure as those introduced by the previous step, which are written
in~(\ref{eq:vec-field-Hr-1}) and~(\ref{eq:def-Hr-1}), respectively.
Indeed, performing the algorithm described in the previous subsection
is advantageous, because the unwanted Hamiltonian terms of degree $0$
and $1$ in the actions get smaller and smaller as $r$ increases
(under the usual KAM hypotheses). This implies that the algorithm is
successful if and only if also the generating functions decrease with
$r$ (recall the equations~(\ref{eq:chi_1}) and~(\ref{eq:chi_2})); this
remark can be easily translated in a numerical test concerning the
construction of the normal form.

The behavior of the sequence of the external forcing frequency vectors
$\big\{\vet{\Omega}^{(r)}\big\}_{r\ge 0}$ deserves a particular
discussion.  Let us recall that the main theorem
in~\cite{Stef-Loc-2012} actually proves the existence of a pair of
objects: an initial frequency vector $\vet{\Omega}^{(0)}$ and a
canonical transformation $\psi^{(\infty)}$ such that the equations of
motion~(\ref{eq:vec-field-H0}) are conjugated to those
in~(\ref{eq:vec-field-Hinfty}), where the Hamiltonian part is in the
Kolmogorov's normal form~(\ref{eq:def-Hinfty}). Thus, the proof scheme
determines $\vet{\Omega}^{(0)}$ {\it a posteriori}, i.e., as a result
of the normalization procedure (actually, here we followed the
approach originally designed in~\cite{Cel-Chi-2009}). Of course, this
is unpractical when we focus on comparisons with numerical results,
because any integrator of the equations of
motion~(\ref{eq:vec-field-H0}) requires that $\vet{\Omega}^{(0)}$ must
be known in advance. This is the main reason why, in
subsection~\ref{sbs:appr-forc-freq} we developed a Newton method based
on the frequency analysis, so to provide a good approximation of the
initial $\vet{\Omega}^{(0)}$, corresponding to the fixed angular
velocity vector $\vet{\omega}$ of the quasi-periodic motion on the
wanted invariant torus. If such a vector $\vet{\Omega}^{(0)}$ would be
perfectly determined, then the algorithm constructing the conjugacy
to~(\ref{eq:vec-field-Hinfty}) (where any external frequency is not
appearing) requires that $\vet{\Omega}^{(r)}\to\vet{0}$ for
$r\to\infty\,$. In the following,  we will test numerically this condition.

\begin{figure}
\centerline{\includegraphics[width=110mm]{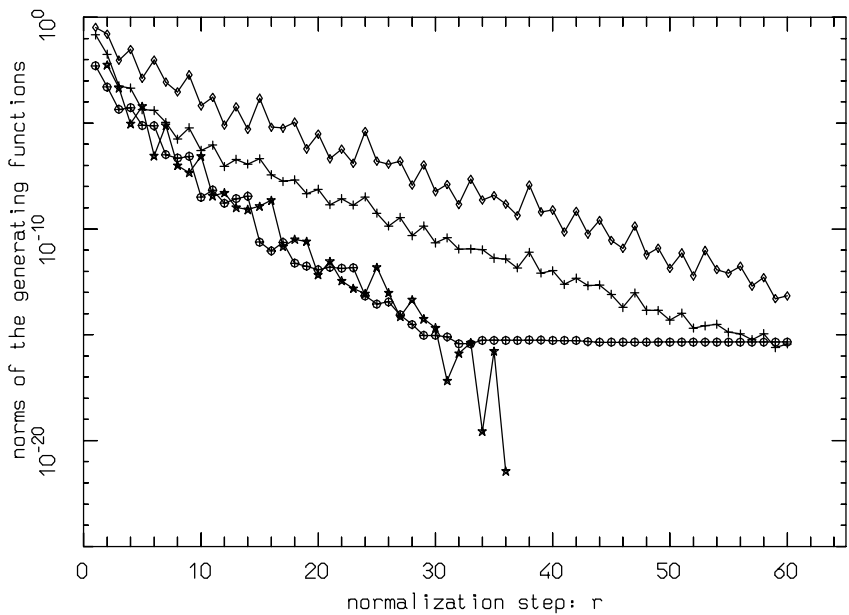}}
\caption{Explicit construction of the Kolmogorov's normal form for the
  dissipative forced pendulum, defined by the
  equations~(\ref{eq:Ham_pendolo})--(\ref{eq:pendolo-dissip-pseudo-Ham}),
  with $\epsilon=0.03$ and $\eta=0.1\,$. Norms of the generating
  functions $X^{(r)}$, $\vet{\csi}^{(r)}$, $\chi_2^{(r)}$ and of the
  external forcing frequency vectors $\vet{\Omega}^{(r)}$ as a
  function of the normalization step $r\,$. Actually, in the figure
  above, their values have been plotted by using the symbols
  $+\,,\>\star\,,\>{\scriptstyle\lozenge}\,,\>{\scriptstyle\oplus}$,
  respectively. See the text for more details.}
\label{fig:numcheck1-torodoro-eps0.03-eta0.1}
\end{figure}

From a practical point of view, let us proceed to a further study of
the equations of motion~(\ref{eq:pendolo-dissip-pseudo-Ham}), where
the Hamiltonian part $H_{\epsilon}$ is given in~(\ref{eq:Ham_pendolo})
and the values of the parameters are fixed so that
\begin{equation}
\epsilon=0.03\ ,
\qquad
\eta=0.1\ ,
\qquad
\Omega=\tilde\Omega^{(5)}=0.3870821721708347\ .
\label{eq:def-parametri-torodoro-eps0.03-eta0.1}
\end{equation}
Let us recall that for this system the frequency analysis results
of subsection~\ref{sbs:appr-forc-freq} clearly show the
existence of an attracting invariant torus, characterized by a
quasi-periodic motion related to the golden frequency
$\omega_1^*=2-\golden=(3-\sqrt{5})/2\,$. In our opinion, the explicit
construction of the normal form related to that torus is rather
challenging, because the 
the perturbing parameter $\epsilon$ is larger than the breakdown threshold value for
the conservative case (i.e., ${\bar\epsilon}_c\simeq 0.0275856$) and at the same
time
it is not so far from the breakdown threshold corresponding to the chosen
friction coefficient $\eta=0.1$ (i.e., $\epsilon_c\simeq 0.0372\,$,
see the table appearing in
Figure~\ref{fig:tab-graf-eta-epsilon-pend-torodoro}).

Of course, it is convenient to reformulate the equations of motion in
the form $\big(\dot{\vet{p}},\dot{\vet{q}}\big)=
\Hamvecfield\left(H^{(0)}\right)-
\eta\big(\vet{p}-\vet{\Omega}^{(0)}\,,\,\vet{0}\big)$ where all the
terms appearing in the expansion of the Hamiltonian $H^{(0)}$ are
determined as in the discussion following formula~(\ref{eq:def-H0}).
In particular, we have
$$
\vet{\omega}=\left(\frac{3-\sqrt{5}}{2}\,,\,1\right)\ ,
\qquad
\vet{\Omega}^{(0)}=
\left(\tilde\Omega^{(5)}-\frac{3-\sqrt{5}}{2}\,,\,0\right)=
(0.0047704825942882482\,,\,0)\ .
$$
Starting from these settings, we explicitly performed $60$~steps of
the normalization procedure described in
subsection~\ref{sbs:Kolm_norm_form}, by using the software package
{\it X$\rho$\'o$\nu o\varsigma$}, that is designed for making computer
algebra, with a special care to its possible applications to Celestial
Mechanics problems (see~\cite{Gio-San-2012} for an introduction to its
main concepts).  Since none of the canonical transformations
prescribed by the algorithm increases the polynomial degree, all the
expansions of the Hamiltonian parts are rather compact, because they
are at most quadratic in the actions as the initial $H^{(0)}$.
Figure~\ref{fig:numcheck1-torodoro-eps0.03-eta0.1} shows
the norms of the generating functions $X^{(r)}$, $\vet{\csi}^{(r)}$
and $\chi_2^{(r)}$ with the index $r\in[1,60]\,$; more precisely, we
have calculated the sum of the absolute values of the coefficients
appearing in~(\ref{eq:sol-omol-per-X}), (\ref{eq:chi_1})
and~(\ref{eq:sol-omol-per-chi2}), respectively. Actually, the
precision of the computation is affected by the truncation rules on
the expansions, that we arranged so to neglect all the terms having a
Fourier harmonic $\vet{k}$ with $l_1$--norm
$|\vet{k}|=\sum_{i=1}^{2}|k_i|>122\,$. In particular, this implies
also that the contributions due to many relevant terms independent
from the angles are not taken into account for $r>30\,$; thus, the
plot of $\big|\vet{\csi}^{(r)}\big|$ has been stopped when the values
corresponding to some indexes $r$ have begun to be unrealistically
small with respect to the previous ones. Let us emphasize that the
geometrical decrease of the generating functions looks quite sharp in
the semi--log scale of
Figure~\ref{fig:numcheck1-torodoro-eps0.03-eta0.1}; this behavior is
in agreement with the analytical estimates on the algorithm
constructing the Kolmogorov's normal form, when it is reformulated
according a {\it classical} formal scheme (see~\cite{Gio-Loc-1997.1}).
Furthermore, also the $l_1$--norm of the external forcing frequency
vector $\vet{\Omega}^{(r)}$ is reported in
Figure~\ref{fig:numcheck1-torodoro-eps0.03-eta0.1}. In this case, the
geometrical decrease is rather sharp until a ``saturation threshold
value'', that is of order $10^{-16}$; for $r>30$ the value of
$\big|\vet{\Omega}^{(r)}\big|$ is approximately constant.  This
unpleasant phenomenon can be easily explained, by taking into account
that the numerical determination of the initial $\vet{\Omega}^{(0)}$
is affected by the unavoidable round-off errors. Such an uncertainity
(due to the application of the frequency analysis numerical method) is
propagated to all the sequence of the external forcing frequency
vectors by the recursive definitions~(\ref{eq:per-hatomegone}) and
(\ref{eq:per-omegone}). Thus, the computed plot of
$r\mapsto\big|\vet{\Omega}^{(r)}\big|$ agrees with the expectation
that $\lim_{r\to\infty}\vet{\Omega}^{(r)}=\vet{0}\,$.  Finally, we can
conclude that Figure~\ref{fig:numcheck1-torodoro-eps0.03-eta0.1} makes
evident that the constructing procedure is converging to the
Kolmogorov's normal form for dissipative systems, which is
characterized by equations~(\ref{eq:vec-field-Hinfty})
and~(\ref{eq:def-Hinfty}).

We now perform another test, checking the accuracy of the conjugacy
canonical transformation $\Kscr^{(r)}$ which is provided after having
carried out the $r$--th normalization step according to the
definition~(\ref{eq:trasf_Kolmogorov}). Some previous works studying
the construction of the Kolmogorov's normal form stressed that it can
be used to integrate the equations of motion on an invariant torus
characterized by a frequency vector $\vet{\omega}\,$. In fact, one can
refer to the following ideal scheme (see, e.g.,~\cite{Loc-Gio-2000}
and~\cite{Gab-Jor-Loc-2005}):
\begin{equation}
\vcenter{\openup1\jot\halign{
 \hbox to 12 ex{\hfil $\displaystyle {#}$\hfil}
&\hbox to 12 ex{\hfil $\displaystyle {#}$\hfil}
&\hbox to 30 ex{\hfil $\displaystyle {#}$\hfil}\cr
\big(\vet{p}(0),\vet{q}(0)\big)
&\build{\longrightarrow}_{}^{{{\displaystyle
\left(\psi^{(\infty)}\right)^{-1}}
\atop \phantom{0}}}
&\left({{\displaystyle \vet{P}(0)=\vet{0}}
\,,\, {\displaystyle \vet{Q}(0)}}\right)
\cr
& &\bigg\downarrow \build{\Phi_{\vet{\omega}\cdot\vet{P}}^{t}}_{}^{}
\cr
\big(\vet{p}(t),\vet{q}(t)\big)
&\build{\longleftarrow}_{}^{{{\displaystyle 
\psi^{(\infty)}} \atop \phantom{0}}}
&\left({{\displaystyle  \vet{P}(t)=\vet{0}}
\,,\, {\displaystyle \vet{Q}(t)=\vet{Q}(0)+\vet{\omega}t}}\right)
\cr
}}
\ \ \,,
\label{semi-analytical_scheme}
\end{equation}
where $\Phi_{\vet{\omega}\cdot\vet{P}}^{t}$ is nothing but the flow
induced by $\vet{\omega}\cdot\vet{P}\,$, that is the only effective
part for the normalized equations of
motion~(\ref{eq:vec-field-Hinfty})--(\ref{eq:def-Hinfty}), when
$\vet{P}=\vet{0}$; moreover, let us recall that
$\psi^{(\infty)}=\lim_{r\to\infty}\Kscr^{(r)}$ is the conjugacy
transformation whose existence is ensured by a KAM--like statement
(under suitable hypotheses). Of course, $\psi^{(\infty)}$ cannot be
explicitly calculated, but this can be done for (a truncated expansion
of) $\Kscr^{(r)}$ with a possibly large value of the index $r\,$.
Thus, it is convenient to limit ourselves to consider a numerical
approximation $t\mapsto\big(\vet{p}(t),\vet{q}(t)\big)$ of the motion
law on an invariant torus characterized by a frequency vector
$\vet{\omega}\,$, so to check if the following relation is satisfied:
\begin{equation}
\Kscr^{(r)}_{i}\big(\vet{p}(t),\vet{q}(t)\big)\simeq 0\ ,
\qquad
\forall\ t\in\reali\ ,
\label{eq:test-action-semi-analytical_scheme}
\end{equation}
where $\Kscr^{(r)}_{i}$ provides the approximately normalized $i$--th
action for $i=1,\,\ldots\,,\,n_1\,$.

\begin{figure}
\centerline{\includegraphics[width=145mm]{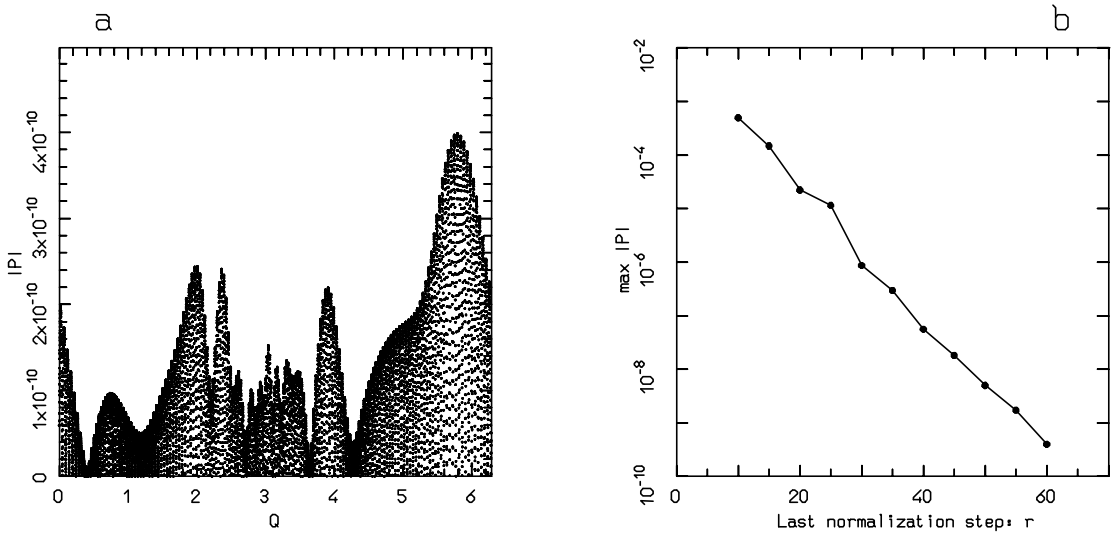}}
\caption{Explicit construction of the Kolmogorov's normal form for the
  dissipative forced pendulum, defined by the
  equations~(\ref{eq:Ham_pendolo})--(\ref{eq:pendolo-dissip-pseudo-Ham}),
  with $\epsilon=0.03$ and $\eta=0.1\,$. In
  Figure~\ref{fig:numcheck2-torodoro-eps0.03-eta0.1}a, $10\,001$ pairs
  of normalized canonical coordinates $(P,Q)$ are plotted for an orbit
  lying on the attracting golden torus; their values are approximately
  computed by applying the inverse of the canonical
  transformation~(\ref{eq:trasf_Kolmogorov}) with $r=60$ to the
  corresponding points, which are generated by the Poincar\'e map of a
  flow $t\mapsto\big(\vet{p}(t),\vet{q}(t)\big)\,$. In
  Figure~\ref{fig:numcheck2-torodoro-eps0.03-eta0.1}b, the maximum of
  the absolute value of the action $P$ is plotted in a semi-log scale
  as a function of the final normalization step $r$ contributing to
  the canonical transformation~(\ref{eq:trasf_Kolmogorov}); $\max|P|$
  is calculated starting from plots similar to
  Figure~\ref{fig:numcheck2-torodoro-eps0.03-eta0.1}a, when
  $r=10,\,15,\,20,\,\ldots\,,\,60\,$.}
\label{fig:numcheck2-torodoro-eps0.03-eta0.1}
\end{figure}

We check formula~(\ref{eq:test-action-semi-analytical_scheme}) with
$r=60$ and, again, in the special case of the dissipative forced
pendulum, that is defined by the
equations~(\ref{eq:pendolo-dissip-pseudo-Ham})
and~(\ref{eq:Ham_pendolo}) with the values of the parameters given
in~(\ref{eq:def-parametri-torodoro-eps0.03-eta0.1}).  In particular,
we consider the motion law $t\mapsto\big(\vet{p}(t),\vet{q}(t)\big)$
on the attracting invariant torus, related to the golden frequency
$\omega_1^*=(3-\sqrt{5})/2\,$.  In
Figure~\ref{fig:numcheck2-torodoro-eps0.03-eta0.1}a, we report the
values of the approximately normalized action
$P=\Kscr^{(60)}_{1}\big(\vet{p}(2j\pi),\vet{q}(2j\pi)\big)$ as a
function of its canonically conjugated angle $Q\simeq
Q(0)+2j\omega_1^*\pi\,$, when $j=0,\,1,\,\ldots\,,\,10\,000\,$.  Let
us stress that the initial value of $Q(0)$ (and the corresponding
$P(0)$) is unrelevant, when we are interested in checking the accuracy
of the normalized canonical coordinates on all the invariant torus,
because it is filled by the quasi-periodic orbit. However, the pair
$(P(0),Q(0))$ is determined after having integrated the equations of
motion for a relaxation time-span, according to the discussion
reported at the end of section~\ref{sec:models}. Let us also remark
that it is natural to neglect completely the second pair of
coordinates; in fact, we recall that the dummy action $p_2$ does not
affect the evolution of all other canonical variables; moreover, since
${\dot q}_2={\dot Q}_2=1\,$, the second angle just describes the
flowing of time. For this reason we have chosen to plot the
approximately normalized first action when $t=2j\pi\,$, as in a
standard Poincar\'e map.
Figure~\ref{fig:numcheck2-torodoro-eps0.03-eta0.1}a highlights that
the action $P$ (making part of the nearly normalized set of canonical
variables) is close to zero for all the considered points generated by
the Poincar\'e map of the flow on the attracting golden torus, in
agreement
with~(\ref{eq:test-action-semi-analytical_scheme}). Actually, the
maximum of $|P|$ is a few orders of magnitude bigger than the
round-off errors threshold, as expected because their accumulation is
unavoidable, while the very large number of computations required by
the expansions are explicitly performed.

We now check the behavior of
formula~(\ref{eq:test-action-semi-analytical_scheme}) as a function of
the final normalization step $r\,$. Since the ideal normalization
transformation is such that
$\psi^{(\infty)}=\lim_{r\to\infty}\Kscr^{(r)}$, then we expect that
\begin{equation}
\lim_{r\to\infty}\sup_{j}\left|
\Kscr^{(r)}_{1}\big(\vet{p}(2j\pi),\vet{q}(2j\pi)\big)\right| =0\ ,
\end{equation}
where $t\mapsto\big(\vet{p}(t),\vet{q}(t)\big)$ denotes again the
motion law on the attracting golden torus. The results of our tests of
the previous formula are illustrated in
Figure~\ref{fig:numcheck2-torodoro-eps0.03-eta0.1}b, where the $\sup$
appearing in the r.h.s. is approximated with the maximum on the index
$j$ ranging in $[0,\,10\,000]\,$. Also in this case, the plot nicely
shows a geometrically decreasing behavior, in agreement with the
expectations (see~\cite{Gio-Loc-1997.1}). This positively ends the
check of the explicit construction of the conjugacy canonical
transformation.

\subsubsection{The basin of attraction of an invariant torus: a semi-analytic lower estimate}\label{sss:rigorous-basin-attraction}
Let us now focus on the dynamics in a region surrounding an invariant
torus, as it is described by {\it normalized coordinates}. This means
that we are going to study the equations of
motion~(\ref{eq:vec-field-Hinfty}), where the Hamiltonian
$H^{(\infty)}$ is of type~(\ref{eq:def-Hinfty}), that is in
Kolmogorov's normal form. Moreover, let us suppose to know some upper
bounds on the size of its terms which depend on the angles and, then,
are at least quadratic with respect to the actions. This allows us to
easily produce some estimates on the basin of attraction, by adapting
a standard technique commonly used in the local theory around an
equilibrium point of ordinary differential equations.

For the sake of simplicity, we summarize the argument
by referring to the Kolmogorov's normal form of
the specific case of a dissipative forced pendulum, where
the pseudo-Hamiltonian equations of motion are the following:
\begin{equation}
\left(\dot P_1,\dot P_2,\dot Q_1,\dot Q_2\right)=
\Hamvecfield\Big(\omega_1P_1+P_2+\Rscr(P_1,Q_1,Q_2)\Big)
-\eta\big(P_1\,,\,0\,,\,0\,,\,0\big)\ ,
\label{eq:pendolo-dissip-pseudo-Ham-Kolm-norm-form}
\end{equation}
where $\Rscr(P_1,Q_1,Q_2)=\Oscr(P_1^2)$ and
$\vet{\omega}=\big(\omega_1,1\big)$ is the frequency vector
characterizing the quasi-periodic motion on the invariant torus
$P_1=0\,$. It is convenient to study the integral form of the first
component of the differential equations
system~(\ref{eq:pendolo-dissip-pseudo-Ham-Kolm-norm-form}), that is
\begin{equation}
P_1(t)=e^{-\eta t}P_1(0)-\int_0^t\diff s\,\bigg[e^{-\eta(t-s)}
\,\frac{\partial\Rscr}{\partial Q_1}\big(P_1(s),Q_1(s),s\big)\bigg]\ .
\label{eq:forma-int-moto-P1}
\end{equation}
Since we supposed to be able to evaluate the size of the terms
depending on the angles, we can assume to know $B\in\reali_+$
such that 
\begin{equation}
\sup_{\vet{Q}\in\toro^2}
\left|\frac{\partial\Rscr}{\partial Q_1}\big(P_1,Q_1,Q_2\big)\right|
\le B P_1^2\ .
\label{diseq:up-bound-deriv-ang}
\end{equation}
Therefore, the following inequality can be immediately deduced
starting from~(\ref{eq:forma-int-moto-P1}):
\begin{equation}
\big|P_1(t)\big|\le e^{-\eta t}\big|P_1(0)\big|+B\int_0^t\diff s\,\Big[e^{-\eta(t-s)}
\,\big(P_1(s)\big)^2\Big]\ .
\label{diseq:forma-int-moto-P1}
\end{equation}
Let us consider a generic initial condition
$\big(P_1(0),\vet{Q}(0)\big)=\big(P_{1;0},\vet{Q}_0\big)$ belonging to
a (not arbitrarily large) set such that $\vet{Q}_0\in\toro^2$ and
\begin{equation}
P_{1;0}\in\Bscr_{\rho}(0)
\quad{\rm with}\quad
\rho\le\frac{\eta-2\mu}{B}\ ,
\label{cond-iniz-P1}
\end{equation}
with $0<\mu<\eta/2\,$. Let us define the positive\footnote{Let us
  remind that, by using~(\ref{diseq:up-bound-deriv-ang}), the velocity
  ${\dot P}_1$ can be easily bounded on the compact set
  $\overline{\Bscr_{(\eta-\mu)/B}(0)}\times\toro^2$; thus,
  $T_{\rho,\mu}$ cannot be equal to $0\,$.}  time $T_{\rho,\mu}$
so that
\begin{equation}
T_{\rho,\mu}=\inf_{{P_{1;0}\in\Bscr_{\rho}(0)}\atop{\vet{Q}_0\in\toro^2}}
\left\{t^\star>0\,:\ \big|P_1\big(t^*;P_{1;0},\vet{Q}_0\big)\big|=
\frac{\eta-\mu}{B}\right\}\ ,
\label{def:tempo-Gronwall}
\end{equation}
where $t\mapsto P_1\big(t;P_{1;0},\vet{Q}_0\big)$ is nothing but the
motion law of the coordinate $P_1$ starting from the initial
conditions
$\big(P_1(0),\vet{Q}(0)\big)=\big(P_{1;0},\vet{Q}_0\big)\,$.  Here, it
is convenient to introduce $\Zscr(t)=\big|P_1(t)\big|e^{\eta
  t}$. Starting from~(\ref{diseq:forma-int-moto-P1}), we can write the
following chain of inequalities:
\begin{equation}
\Zscr(t)\le \big|P_1(0)\big|+
(\eta-\mu)\int_0^t\diff s\,\Zscr(s)
\le \big|P_1(0)\big|e^{(\eta-\mu)t}
\qquad
\forall\ 0\le t\le T_{\rho,\mu}
\ ,
\label{diseq:stima-Gronwall}
\end{equation}
where we used the well known Gronwall's lemma.  The previous formula
can be rephrased for the law motion of the variable $P_1\,$, so that
$\big|P_1(t)\big|\le \big|P_1(0)\big|e^{-\mu t}$ $\forall\ 0\le t\le
T_{\rho,\mu}\,$. Therefore, also $P_1(T_{\rho,\mu})\in\Bscr_{\rho}(0)$
and then we can extend the procedure also to the time intervals
$\big[T_{\rho,\mu}\,,\,2T_{\rho,\mu}\big]\,$,
$\big[2T_{\rho,\mu}\,,\,3T_{\rho,\mu}\big]$ and so on. This allows us
to justify the exponential estimate $\big|P_1(t)\big|\le
\big|P_1(0)\big|e^{-\mu t}$ $\forall\ t\ge 0\,$. Since the parameter
$\mu$ can be made arbitrarily small, we can finally conclude that
\begin{equation}
\lim_{t\to\infty}P_1\big(t;P_{1;0},\vet{Q}_0\big)=0
\qquad
\forall\ \big(P_{1;0}\,,\,\vet{Q}_0\big)\in\Bscr_{\eta/B}(0)\times\toro^2\ .
\label{limite-toro-attrattore}
\end{equation}
Let us emphasize that this same approach can be rather trivially
extended, so to obtain the same result for higher dimensions
pseudo-Hamiltonian equations of motion, where the initial Kolmogorov's
normal form Hamiltonian $H^{(\infty)}$ is of
type~(\ref{eq:def-Hinfty}).

\begin{figure}
\centerline{\includegraphics[width=145mm]{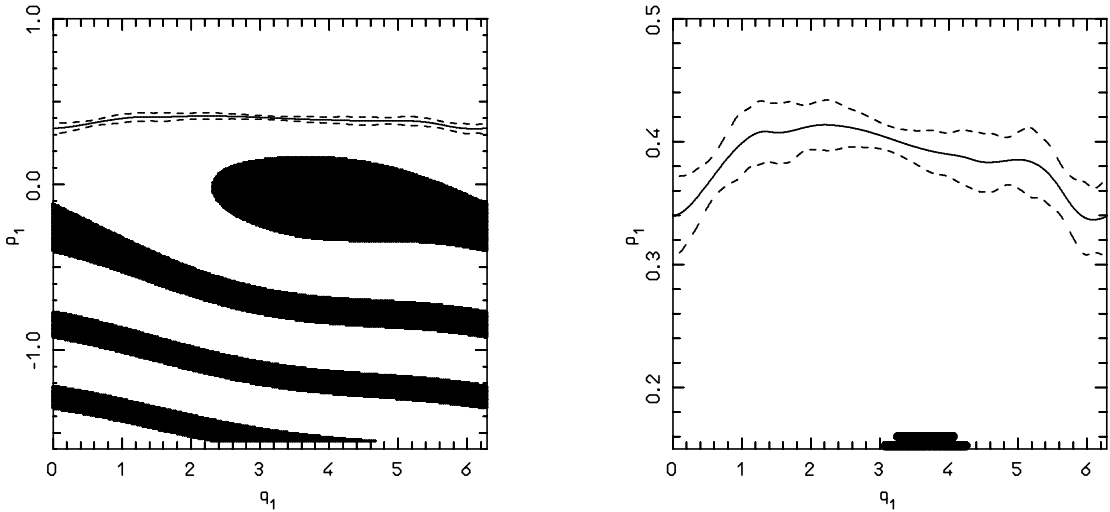}}
\caption{Attractors and their basins for the dissipative forced
  pendulum, defined by the
  equations~(\ref{eq:Ham_pendolo})--(\ref{eq:pendolo-dissip-pseudo-Ham}),
  in the case with $\epsilon=0.028\,$, $\eta=0.05$ and
  $\Omega=0.3867364938443934\,$: study of the Poincar\'e map
  corresponding to the section
  $q_2=t=0\,$. Figure~\ref{fig:bac_attr_toro_+_orb_per}b is nothing
  but an enlargement of~\ref{fig:bac_attr_toro_+_orb_per}a. In both
  panels, the solid curve locates the (section of the) invariant
  torus, related to the golden frequency $(3-\sqrt{5})/2\,$, while the
  region between the dashed lines is certainly included in the basin
  of attraction of that torus, according to our semi-analytic
  evaluations. The black region describes another basin of attraction
  of a periodic orbit.}
\label{fig:bac_attr_toro_+_orb_per}
\end{figure}

Formula~(\ref{limite-toro-attrattore}) can be directly applied to the
dissipative forced pendulum, so to locate a subset of the basin of
attraction of the invariant torus. In fact, let us consider an initial
condition
\begin{equation}
\big(p_{1;0},\vet{q}_0\big)\in
\left(\psi^{(\infty)}\right)^{-1}\Big(\Bscr_{\eta/B}(0)\times\toro^2\Big)
\label{sottoinsieme-bacino-di-attrazione}
\end{equation}
where $\psi^{(\infty)}$ is the change of
coordinates\footnote{Actually, here we avoid to consider the effect of
  the change of coordinates on the dummy actions $p_2$ and $P_2\,$,
  because they do not affect the evolution of all other variables. In
  the present subsection, this abuse of notation will be made also for
  the transformation approximating $\psi^{(\infty)}$.} appearing in
the ideal scheme~(\ref{semi-analytical_scheme}); this means that it
brings the pseudo-Hamiltonian equation of motion to the Kolmogorov's
normal form. Therefore, the corresponding motion law
$t\mapsto(p_1,\vet{q})\big(t;p_{1;0},\vet{q}_0\big)$ tends (in an
exponentially fast way) to the invariant tous, in view
of~(\ref{limite-toro-attrattore}) and because $\psi^{(\infty)}$ is
canonical (see~\cite{Stef-Loc-2012}). For practical purposes, the
change of coordinates $\psi^{(\infty)}$ must be replaced, of course,
by $\Kscr^{(r)}$, that is defined in~(\ref{eq:trasf_Kolmogorov}) and
is given by the composition of the transformations related to the
first $r$ normalization steps.
Figure~\ref{fig:bac_attr_toro_+_orb_per} represents the intersection
of the initial conditions~(\ref{sottoinsieme-bacino-di-attrazione})
with the plane $q_2=Q_2=t=0\,$; they are located in the region between
the two dashed curves.  Actually, those curves have been drawn by
plotting the inverse image of the two rings
$\{P_1=\pm\eta/B\,,\ Q_1\in\toro\,,\ Q_2=0\}$ with respect to the map
$\Kscr^{(60)}$, in the case of the dissipative forced pendulum
equations~(\ref{eq:Ham_pendolo})--(\ref{eq:pendolo-dissip-pseudo-Ham}),
when the values of the parameters are fixed so that
$$
\epsilon=0.028\ ,
\quad
\eta=0.05\ ,
\quad
\Omega=0.3867364938443934\ .
$$
In that case, there are two attractors: the invariant torus
corresponding to the vector frequency
$\vet{\omega}=\big((3-\sqrt{5})/{2}\,,\,1\big)$ and a periodic orbit;
their corresponding sections with the plane $q_2=0$ are located by a
solid line, that can be seen in
Figure~\ref{fig:bac_attr_toro_+_orb_per}, and a fixed point having
coordinates $\simeq(3.923867\,,\,-0.021613)$, respectively. By the
way, let us recall that other examples of dissipative systems showing
the coexistence of more than one attractor are described
in~\cite{Cel-Chi-2008}. The change of coordinates $\Kscr^{(60)}$, that
is a good approximation of the ideal normalizing transformation
$\psi^{(\infty)}$, is calculated by following the detailed discussion
of subsections~\ref{sbs:Kolm_norm_form}
and~\ref{sss:check-Kolm-norm-form}; moreover, $B$ is evaluated so to
satisfy inequality~(\ref{diseq:up-bound-deriv-ang}), when the
remainder $\Rscr$ is replaced by the finite sum of the quadratic terms
belonging to the calculated truncation of the Hamiltonian $H^{(60)}$.
Of course, the drawing of the section of the initial
conditions~(\ref{sottoinsieme-bacino-di-attrazione}) is made in a
numerical way; nevertheless, we emphasize that the procedure could be
made completely rigorous, so to determine a set that is certainly
included in the basin of attraction of the invariant torus, by
implementing interval arithmetics and providing analytic estimates of
all the truncated terms (see, e.g.,~\cite{Cel-Gio-Loc-2000}). The
effects of such a computer-assisted procedure are expected to be
completely unrelevant in a plot like that of
Figure~\ref{fig:bac_attr_toro_+_orb_per}.  In order to check the
effectiveness of our procedure, the basin of attraction of the
periodic orbit is drawn in black in
Figure~\ref{fig:bac_attr_toro_+_orb_per}. Of course, the area provided
by our estimates cannot cover a rather significant part of the basin
of attraction of the invariant torus, that is actually expected to be
infinitely big, because it is the complementary of the black region
appearing in Figures~\ref{fig:bac_attr_toro_+_orb_per}a
and~\ref{fig:bac_attr_toro_+_orb_per}b.  Nevertheless, we think that
our approach can be useful to develop computational
methods also in dissipative systems with a larger number of degree of
freedom. Indeed, it could help to locate an initial thin region of the
basin of attraction surrounding the possibly complicated shape of an
invariant torus; therefore, the computational strategy could include
suitable numerical integrations backwards in time, so to reconstruct
all the parts of the phase space that are potentially belonging to the
basin; moreover, some final forward integrations (taking care of the
propagation of the errors) could validate the location of most of the
basin of attraction.

\section{Conclusions}\label{sec:conclu}
Since the sixities and during a couple of decades, KAM theorem was
commonly considered to be a very elegant mathematical result, but
substantially irrelevant for real problems in physics, because the
hypothesis on the smallness of the perturbation was (and still is)
extremely restrictive (see~\cite{Henon-66}).  Indeed, several articles
appeared since the eighties actually showed that KAM theory can be
effectively applied to realistic models, provided that it is
complemented with suitable computational techniques (see,
e.g.,~\cite{Cel-Chi-2007}). In our opinion, this work adds a few new
arguments to such a more modern point of view.

KAM theory provides the natural framework to define the frequency
analysis method, which discriminates between quasi-periodic motions
and chaotic ones in Hamiltonian systems (\cite{Laskar-99}
and~\cite{Laskar-2005}); this allows a global understanding of the
dynamics.  In section~\ref{sec:method}, that computational approach
has been adapted to a special class of dissipative systems with
friction terms, that are linear and homogeneous with respect to the
actions. The method is based on a clear interpretation of the results,
that can be nicely visualized. Moreover, our evaluation of the
breakdown threshold for invariant tori in the dissipative standard map
is in agreement with some existing results in literature
(see~\cite{Cal-Cel-2010}). Actually, our method is less precise than
that based on the computation of the Sobolev norms, also because our
approach, which is extremely visual, is hard to made automatic; this
strongly limits the performances when a great accuracy is
required. This is also the reason why, in our opinion, the numerical
results given in section~\ref{sec:results-diss-forc-pend-model} (about
the breakdown threshold of invariant tori for the dissipative forced
pendulum) show just partially the expected behavior.

In section~\ref{sec:kolmog}, the explicit algorithm constructing the
Kolmogorov's normal form has been adapted so to cover also the case of
dissipative systems having a pseudo-Hamiltonian framework. Moreover,
such a reformulation has been successfully applied to the dissipative
forced pendulum model: by a code implementing algebraic manipulations
on a computer, the shape of an attracting invariant ``golden'' torus
(related to some specific values of the parameters) has been carefully
reconstructed. We emphasize that such a result was expected, but it is
not trivial, because the constructive algorithm producing the normal
form for the dissipative case is significantly different with respect
to that traditionally used in KAM theory for Hamiltonian systems. Let
us recall that the good system of equations (for which the ``golden''
torus is invariant and attracting) has been settled by using the
frequency analysis, so to determine {\it a priori} the numerical value
of one of the parameters (namely, the external frequency
$\Omega=\tilde\Omega^{(5)}$
in~(\ref{eq:def-parametri-torodoro-eps0.03-eta0.1})). In our opinion,
this fact validates our implementations of both the frequency analysis
and the construction of the Kolmogorov's normal form, because two so
different techniques provide results that are in agreement between
them. As a further natural application of a method based on a normal
form, in the last subsection~\ref{sss:rigorous-basin-attraction}, the
contracting dynamics in a neighborhood of an invariant torus is
estimated by using the Gronwall's inequality. This has allowed us to
show that such an open set is certainly included in the basin of
attraction of that invariant torus.

\subsection*{Acknowledgments}
A.~Celletti encouraged us to study the particular class of dissipative
systems considered in the present paper. A.~Giorgilli allowed us to
use the computer algebra package {\it X$\rho$\'o$\nu o\varsigma$},
initially written by himself with the late (and relevant)
contribution of M.~Sansottera. A.~Noullez suggested us how to improve
our numerical code doing frequency map analysis.  We are deeply
indebted with all of them.

\end{document}